\documentclass[journal]{IEEEtran}
\usepackage{cite}
\usepackage{amsmath,amssymb,amsfonts}
\usepackage{graphicx}
\usepackage{textcomp}
\usepackage[table]{xcolor}
\newcommand{\newtext}[1]{#1}
\usepackage{tcolorbox}
\usepackage{booktabs} 
\usepackage{makecell, cellspace}
\usepackage{caption}
\usepackage{physics}
\usepackage{multirow}
\usepackage{algorithm}
\usepackage{algpseudocode}
\usepackage{mathtools}
\usepackage{amssymb}
\usepackage{pifont}
\usepackage{hyperref}
\definecolor{customred}{HTML}{E77818}
\definecolor{customblue}{HTML}{1F77B4}
\hypersetup{
    colorlinks=true,      
    linkcolor=customblue,        
    citecolor=customblue,       
    urlcolor=customblue         
}
\usepackage{aliascnt}
\usepackage{amsthm} 
\newcommand{\cmark}{\textcolor{customblue}{\ding{51}}}%
\newcommand{\xmark}{\textcolor{customred}{\ding{55}}}%
\newcommand{\thetavec}{\pmb{\theta}}

\def\BibTeX{{\rm B\kern-.05em{\sc i\kern-.025em b}\kern-.08em
    T\kern-.1667em\lower.7ex\hbox{E}\kern-.125emX}}

\newcommand{\bt}[1]{\mbox{$\bf #1$}}
\def\l{\left(}
\def\r{\right)}
\newcommand{\img}{\bt x}
\newcommand{\cimg}{\hat{\bt x}}
\newcommand{\pixres}{\cimg(\thetavec) - \img}
\newcommand{\pixresi}{\cimg_i(\theta_i) - \img_i}
\newcommand{\jaco}{\bt J_{\sf f}}
\newcommand{\jacos}{\bt J_{\mathbf{s}}}

\newcommand{\numpix}{n_p}
\newcommand{\numfeat}{n_{\sf f}}
\newcommand{\nummb}{n_b}

\newcommand{\numrma}{n_s}

\DeclareMathOperator*{\argmin}{arg\,min}

\newcommand{\fpnfe}{\texttt{FPN-FE}{}}
\newcommand{\rpnfefif}{\texttt{RPN-FE(50){}}}
\newcommand{\rpnfehun}{\texttt{RPN-FE(101)}{}}
\newcommand{\yolofe}{\texttt{YOLO-FE{}}}

\newtheorem{theorem}{Theorem}[section]
\newaliascnt{proposition}{theorem}

\newtheorem{proposition}[proposition]{Proposition}

\aliascntresetthe{proposition}

\newtheorem{lemma}[theorem]{Lemma}     

\begin{document}
\title{Image Coding for Machines via Feature-Preserving Rate-Distortion Optimization\\

\author{
\IEEEauthorblockN{Samuel Fernández-Menduiña, \textit{Student Member, IEEE},
 Eduardo Pavez, \textit{Member, IEEE},  \\ and Antonio Ortega, \textit{Fellow, IEEE}}

\thanks{This work was funded in part by the Fulbright Comission in Spain.}
\thanks{S. Fernández-Menduiña, E. Pavez, and A. Ortega are with the Ming Hsieh Department of Electrical and Computer Engineering, University of Southern California, Los Angeles, California, 90089, United States (email: samuelf9@usc.edu; pavezcar@usc.edu; aortega@usc.edu).}
\thanks{Manuscript received Month XX, 20XX; revised Month XX, 20XX.}}
}

\maketitle

\begin{abstract}
Many images and videos are primarily processed by computer vision algorithms, involving only occasional human inspection. When this content requires compression before processing, e.g., in distributed applications, coding methods must optimize for both visual quality and downstream task performance. We first show \newtext{theoretically} that an approach to reduce the effect of compression for a given task loss is to perform rate-distortion optimization (RDO) using the distance between features, obtained from the original and the decoded images,  as a distortion metric. However, optimizing directly such a rate-distortion \newtext{objective is computationally impractical because it requires iteratively encoding and decoding the entire image–plus feature evaluation–for each possible coding configuration}. We address this problem by simplifying the RDO formulation to make the distortion term computable using block-based encoders. We first apply Taylor's expansion to the feature extractor, recasting the feature distance as a quadratic metric \newtext{involving} the Jacobian matrix of the neural network. Then, we replace the linearized metric with a block-wise approximation, which we call input-dependent squared error (IDSE). 
\newtext{To make the metric computable}, we approximate IDSE using \newtext{sketches of the} Jacobian. The resulting loss can be evaluated block-wise in the transform domain and combined with the sum of squared errors (SSE) to address both visual quality and computer vision performance. Simulations with \newtext{AVC and HEVC} across multiple feature extractors and downstream networks show up to \newtext{17}\% bit-rate savings for the same task accuracy compared to RDO based on SSE, with no decoder complexity overhead and  \newtext{a small (7.86\%)} encoder complexity increase.
\end{abstract}

\section{Introduction}
\begin{figure}[t]
    \centering
    \includegraphics[width=\linewidth]{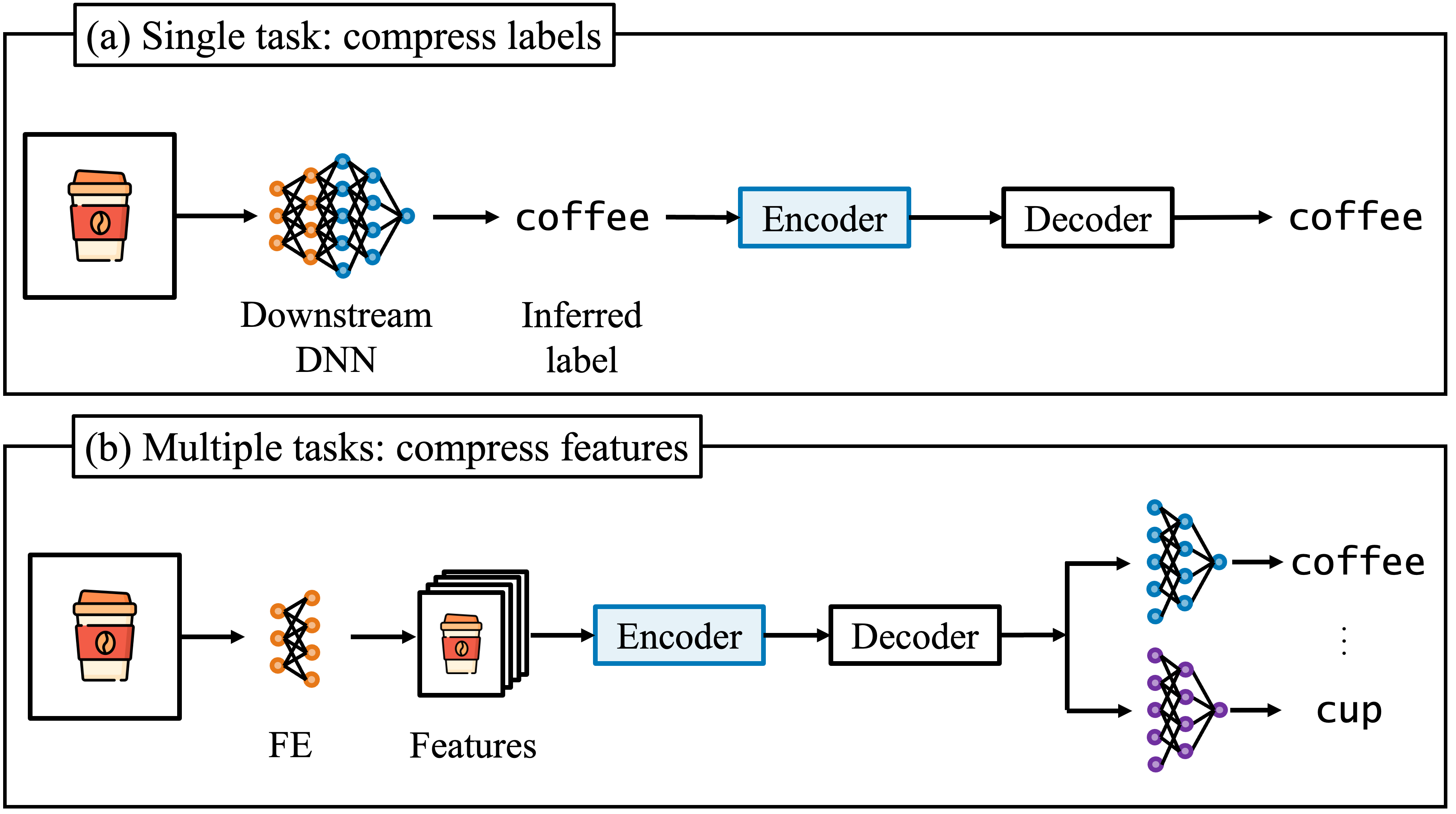}

    \vspace{0.25em}
    \includegraphics[width=\linewidth]{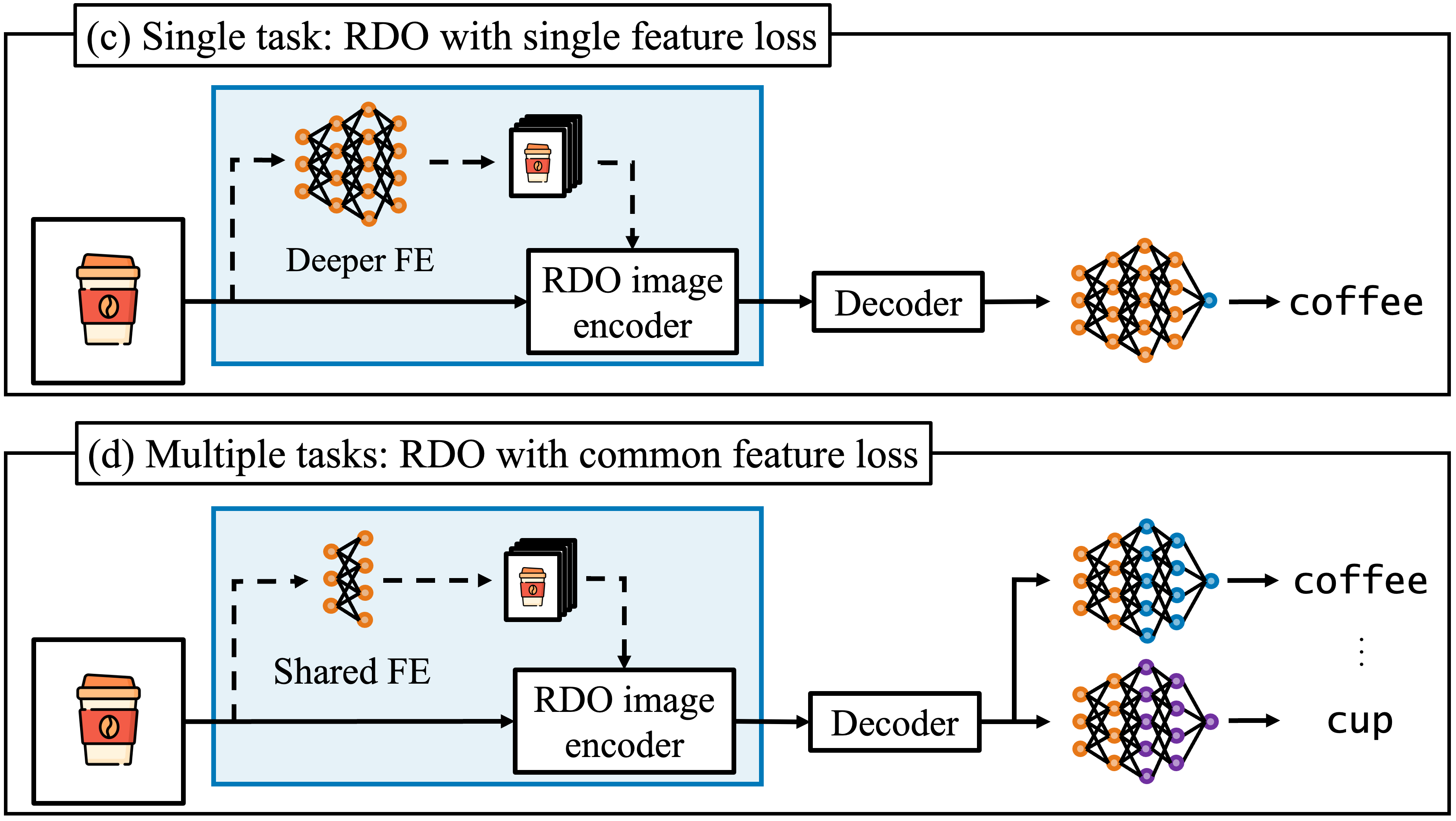}
    \caption{Examples of CM methods when image transmission is (a-b) not needed and (c-d) needed. FE = feature extractor.}
    \label{fig:taxonomy_methods}
\end{figure}

Many images and videos are now primarily consumed by machine learning systems to perform pattern recognition tasks. \newtext{When compression is needed before algorithmic processing, coding artifacts may impact computer vision (CV) performance. To address this problem, some methods propose to code images in ways that specifically mitigate the impact of errors on subsequent machine learning tasks}, a framework known as \emph{coding for machines} (CM) \cite{choi_scalable_2022, zhang_call_2022, ascenso_jpeg_2023}. While similar ideas were explored for classical learning methods \cite{ortega_compression_2000}, advances in deep neural networks (DNNs) \cite{lecun_deep_2015} applied to CV have sparked renewed interest \cite{choi_deep_2018, choi_high_2018, le_learned_2021}. 
\newtext{Two different coding strategies are possible depending on the task requirements, what is known about the task at the encoder, and the power constraints of the system \cite{bajic2025rate, compressai_vision}: 1) running either all or a part of the machine learning model in the sensing device and compressing and transmitting the outputs (i.e., either compressing features or labels), and 2) encoding and transmitting the original image and then running the machine learning model on the decoded image. 
For the first approach, which is suitable when reconstructing the image at the receiver is not required,} algorithms based on the information bottleneck method \cite{tishby1999information} are sufficient. For instance, for single-task classification problems, encoding the inferred labels–\newtext{also known as local inference–}is optimal \cite{dubois_lossy_2021} (\autoref{fig:taxonomy_methods}(a)). For families of related CV tasks, it can be more efficient to compress \emph{feature vectors}, e.g.,  the outputs of the earlier layers of a DNN \cite{hossain2023flexible} (\autoref{fig:taxonomy_methods}(b)). 
Similarly, when the target tasks are unknown, features trained for invariance–e.g., using self-supervised learning (SSL) \cite{bardes2021vicreg}–can be extracted, compressed, and transmitted \cite{dubois_lossy_2021, zhihao2024}. 

We consider instead applications that require \newtext{the reconstruction of the image at the receiver, so the task can be run on the decoded image, also known as remote inference} \cite{choi_scalable_2022}, enabling additional human supervision for the task (cf.~\autoref{fig:taxonomy_methods}(c)–(d)). 
We focus on distributed communication settings with power and bandwidth constraints. Since images have to be transmitted anyway, it is often preferable to perform the CV task remotely on the decompressed image \cite{choi_high_2018} (\autoref{fig:taxonomy_methods}(c)), which is also more efficient given the complexity of running a full-fledged DNN at the sensing system \cite{zhang_call_2022}. This approach is particularly advantageous when running several CV tasks on the same image, as it avoids executing multiple DNNs on the edge device (\autoref{fig:taxonomy_methods}(d)).
Examples of this setup are object detection/instance segmentation in video surveillance, traffic monitoring, or autonomous navigation \cite{jiang_adaptive_2023}.


For our scenarios of interest (\autoref{fig:taxonomy_methods}(c)-(d)), the encoder should be optimized to preserve \textit{both} CV task \newtext{accuracy} and visual quality \cite{ortega_compression_2000}. 
Thus, distortion metrics conventionally used for rate-distortion optimization (RDO) \cite{ortega_rate-distortion_1998, sullivan_rate_1998}, such as the sum of squared errors (SSE), must be complemented or replaced by task-specific losses.  
\newtext{Our main goal in this paper is to provide a general CM framework that can be used to re-purpose traditional codecs–with  SSE-based RDO–to account \newtext{for the effect of lossy coding on the machine learning task(s)}.} 

Based on heuristic arguments, prior work proposed the distance between features obtained from the original and the decoded images (feature distance, FD) \cite{fischer_video_2020, fernandez2024feature} as a distortion metric that can be incorporated into RDO to account for CV performance. 
Since features are extracted from full images but modern codecs can control bit allocation at the block level, \cite{fischer_video_2020} proposes to compute FD using the features extracted from the decoded image for each 
block-wise coding configuration. While allowing RDO with a distortion metric relevant to CV, this setup requires an iterative workflow of encoding,
decoding, and feature extraction for each coding option, which is computationally impractical. 
\cite{fischer_video_2020} addresses this problem by computing block-wise shallow features, which limits performance since the target tasks are completed with deep features extracted from the whole image (cf.~\autoref{sec:blockfd}).

\begin{figure}[t]
    \centering
    \includegraphics[width=\linewidth]{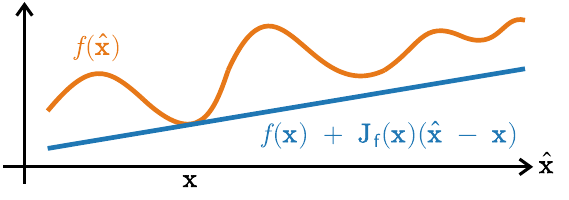}
    \caption{Linear approximation of the features of the compressed image via Taylor's expansion around the input image.}
    \label{fig:taylor_expansion}
\end{figure}

\begin{figure*}[t]
    \centering
    \includegraphics[width=\linewidth]{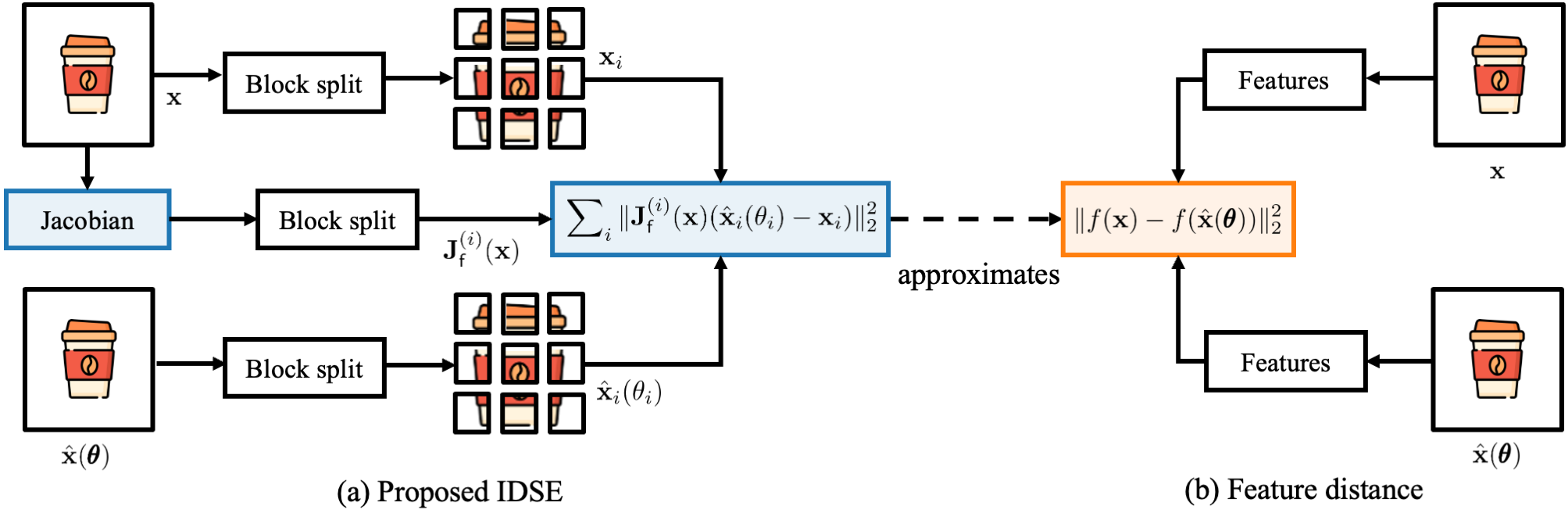}
    \caption{\newtext{Computation of (a) the proposed IDSE and (b) the feature distance (FD). While evaluating FD requires the complete image, IDSE approximates the FD and can be evaluated block-wise, allowing for block-level bit-allocation via RDO.}}
    \label{fig:diagram_idse}
\end{figure*}

\newtext{In this paper, we justify theoretically the use of FD in CM setups and propose a practical framework, which is optimal under high bit-rate assumptions, to account for FD during RDO with better complexity and coding efficiency than existing methods. More precisely,} we show theoretically that minimizing FD can preserve task performance  This analysis can also guide the choice of feature extractor; for instance, when multiple tasks can be addressed with DNNs sharing common earlier layers, as in transfer learning \cite{jain2023data}, using these common layers as a feature extractor can preserve performance in all transferred tasks (\autoref{fig:taxonomy_methods}(d)). To make FD practical for block-level RDO, we first apply Taylor's expansion to recast FD as a quadratic loss involving the Jacobian matrix of the feature extractor with respect to the input image (\autoref{fig:taylor_expansion}). \newtext{The squared norm of the rows of this Jacobian matrix can be seen as an importance map, weighting each pixel based on its relevance for the target tasks.} Then, we replace this linearized metric with a block-wise approximation (\newtext{cf.~\autoref{fig:diagram_idse}}), which we call \emph{input-dependent squared error} (IDSE). We use sketching techniques \cite{achlioptas_database_2003} to avoid computing the entire Jacobian matrix. IDSE can be evaluated block-wise in the transform domain; by combining it with an SSE term, the codec can optimize for both perceptual quality and CV tasks while remaining compatible with standard-compliant decoders.

Results using \newtext{both AVC \cite{wiegand_overview_2003, sullivan2006overview} and HEVC \cite{sullivan_overview_2012} codecs} show that using RDO with IDSE provides up to \newtext{$17$\%}  bit-rate savings with respect to RDO with SSE while preserving the same accuracy for object detection/instance segmentation tasks in the COCO 2017 validation set \cite{lin_microsoft_2014} and the PennFudan dataset \cite{wang_object_2007}. 
\newtext{Although we conduct our experiments with AVC and HEVC,} feature-preserving RDO can be used in systems with more coding options, such as VVC \cite{bross2021overview} (cf.~\autoref{sec:conclusion}).

In this paper, we extend our prior work on feature-preserving RDO  \cite{fernandez2024feature} with:
\begin{itemize}
    \item \emph{theoretical foundations}, including justifications for minimizing FD as a proxy to preserve task loss (\autoref{sec:feat_extract}), \newtext{tighter convergence bounds for the Taylor expansion (\autoref{sec:taylor})}, and \newtext{better arguments for the block-wise approximation resulting in IDSE} (\autoref{sec:blockwise_localization}), 
    \item \emph{input-adaptive methods}, including the selection of the regularization parameter controlling the IDSE-SSE trade-off (\autoref{sec:reg}) and the Lagrangian (\autoref{sec:lag_sel}),
    \item \emph{extended experiments}, including feature extractors of various depths (\autoref{sec:large_scale}) and architectural complexities (\autoref{sec:diff_models}), \newtext{an analysis of the trade-off between visual quality and downstream DNN performance, as well as tests with a mismatch between the network used for IDSE-based feature preservation and the downstream DNN (\autoref{sec:SSL_comp}), and
    \item \emph{practical evaluations} with multiple codecs (\autoref{sec:diff_codecs})}.
\end{itemize}


\subsection{Related work}
\subsubsection{Neural compression} 
Both generative \cite{balle_nonlinear_2020} and input-dependent \cite{ladune2023cool} neural compression methods \cite{balle_nonlinear_2020} can be trained end-to-end to optimize a loss that includes both SSE and a CV downstream task \cite{ascenso_jpeg_2023, li2024human}. However, computational cost limits their deployment in power-constrained platforms \cite{balle2024good}: generative methods require millions of multiply and add operations per pixel at both the encoder and the decoder \cite{guleryuz_sandwiched_2021}, while input-dependent encoders \cite{ladune2023cool} remain too complex for the lower-end devices used in distributed applications.

\subsubsection{Modified traditional codecs} 
Several methods have been proposed to modify conventional codecs, which are less computationally demanding than learned approaches. 
Ahonen et al.\,\cite{ahonen2021learned} enhance decompressed images using learned filters, which results in better performance on the task but cannot be used to optimize bit allocation. Other approaches allow the encoder to be modified. For example, \cite{luo2020rate} modifies JPEG quantization tables using a proxy codec, which cannot be easily extended to modern video compression systems 
with intra/inter-prediction. 
Alternatively, \cite{choi_high_2018} chooses quantization steps at the block level but does not allow optimizing other codec parameters such as block partitioning or prediction mode. 
All these approaches fall short in optimizing compression efficiency relative to RDO-based methods such as \cite{fischer_video_2020}.

\begin{figure}[t]
    \centering
    \includegraphics[width=0.49\linewidth]{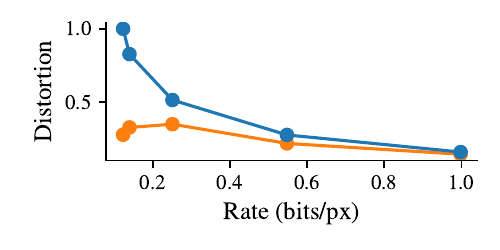}
    \includegraphics[width=0.49\linewidth]{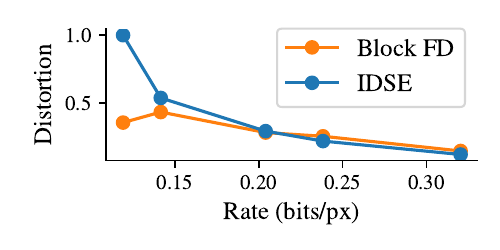}
    \caption{RD curves using SSE-RDO AVC for block FD and IDSE with a feature pyramid network (FPN) \cite{lin_feature_2017} as feature extractor, for blocks of $32\times 32$ pixels. Neural network non-linearities can make block FD concave or non-monotonic with the rate, reducing the number of possible operating points. IDSE is quadratic by design and has monotonic behavior.}
    \label{fig:examples_losses}
\end{figure}

\begin{table}[t]
\setlength{\tabcolsep}{6pt} 
\renewcommand{\arraystretch}{1.1}
\centering
\begin{tabular}{@{}lccccc}
\hline
\multicolumn{1}{@{}l}{\textbf{Distortion}} & \thead{\textbf{Block}\\ \textbf{RDO}} & \thead{\textbf{Trans.}\\\textbf{RDO}} &  \textbf{Quad.} & \textbf{Global} & \textbf{Target} \\
\hline
FD & \xmark & \xmark & \xmark & \cmark & Mach. \\
Block FD \cite{fischer_video_2020} & \cmark & \xmark & \xmark & \xmark & Mach. \\
SSE & \cmark & \cmark & \cmark & \xmark & Humans \\
\textbf{IDSE (ours)} & \cmark & \cmark & \cmark & \cmark & Humans/Mach.\\ 
\bottomrule \\
\end{tabular}
\caption{RDO methods in CM. \newtext{``Mach.''~stands for machines.} Some approaches use quadratic metrics (\textbf{Quad.}), can be applied in the transform domain (\textbf{Trans.}), and use features extracted from the whole image (\textbf{Global}). IDSE \newtext{can reflect both visual and task quality while decomposing as the sum of block-level distortions in the transform domain}.}
\label{tab:rate_dist_optimization}
\end{table}

\subsubsection{RDO based on FD}
\label{sec:blockfd}
Closest to our approach, \cite{fischer_video_2020} achieves block-level RDO  
by evaluating the feature extractor block-wise (block FD).  
Nonetheless, extracting features independently for each block overlooks relationships between blocks and requires using relatively shallow features. As a result, \cite{fischer_video_2020} cannot directly evaluate the final importance of each block for a target task, which is determined in deeper layers.
Moreover, DNN non-linearities often lead to concave or non-monotonic rate-distortion (RD) landscapes, so increasing rates may no longer reduce FD. Thus, the RD trade-off becomes harder to navigate. For instance, only a subset of low/high rate operating points may be reachable (\autoref{fig:examples_losses}), leading to reconstructions with a large SSE even for high rates, which impacts visual quality. Block FD may also become computationally intensive \cite{gou_fast_2023} since it requires pixel-domain evaluation of the DNN for each RDO candidate.

In contrast to \cite{fischer_video_2020}, which computes the FD explicitly for each block 
\textit{using the output of shallow layers} as approximately-local features, our IDSE-based approach uses \textit{backpropagation} via autodiff \cite{paszke_automatic_2017} to determine the importance of each pixel (per-pixel importance map) given \textit{any neural feature extractor} (shallow or deep).
Unlike FD and block FD, IDSE is guaranteed to be quadratic with the pixel-wise error, making the RD curves convex and monotonic. 
Similar to SSE, RDO with IDSE can be performed block-wise in the transform domain and, by relying on the features obtained from the whole image, it can account for the final importance of each block for the target computer vision tasks, which \cite{fischer_video_2020}  cannot do (cf.~\autoref{tab:rate_dist_optimization} for a comparison).

\subsection{Organization and notation}
\textbf{Organization.} \newtext{This paper is organized as follows. \autoref{sec:preliminaries} reviews the classical RDO available in existing codecs} and \autoref{sec:feature_rdo} \newtext{explains our main contribution}, feature-preserving RDO. In \autoref{sec:integration}, we review how to integrate IDSE in a modern block-based codec. \newtext{We include experiments in \autoref{sec:exper}  and provide conclusions and discuss future research in \autoref{sec:conclusion}.}

\textbf{Notation} (cf.~\autoref{tab:notations}). Uppercase bold letters, such as $\bt A$, denote
matrices. Lowercase bold letters, such as $\bt a$, denote vectors.
The $n$th entry of $\bt a$ is $a_n$, and the $(i, j)$th entry of $\bt A$ is $A_{ij}$. Regular letters denote
scalar values.

\begin{table}[t]
    \centering
    \setlength{\tabcolsep}{8pt} 
    \renewcommand{\arraystretch}{1.2}
    \begin{tabular}{@{}ll}
        \toprule
        \textbf{Symbol} & \textbf{Description} \\
        \midrule
        $n_b$ & Number of blocks in the input \\     $n_{\sf f}$ & Dimensionality of feature space \\
        $n_p$ & Number of pixels in the input \\
        $n_{pb}$ & Number of pixels in a block \\        
        $n_r$ & Number of RDO candidates per block \\
        $n_s$ & Feature dimensionality after sketching \\        
        $n_t$ & Number of downstream tasks \\
        \midrule
        $\img\in\mathbb{R}^{\numpix}$ & Codec input \\
        $\img_i\in\mathbb{R}^{n_{pb}}$ & $i$th block of the input \\        $\cimg(\boldsymbol{\theta})\in\mathbb{R}^{\numpix}$ & Compressed version of $\img$ with parameters $\boldsymbol{\theta}$ \\
$\cimg_i(\theta_i)\in\mathbb{R}^{n_{pb}}$ & $i$th block of the compressed image \\        
        \midrule        
        $f(\img)\in\mathbb{R}^{\numfeat}$ & Features extracted from $\img$\\
        $\jaco(\img)\in \mathbb{R}^{\numfeat\times\numpix}$ & Jacobian of $f(\cdot)$ evaluated at $\img$ \\
        $\jaco^{(i)}(\img)\in \mathbb{R}^{\numfeat\times n_{pb}}$ & Columns of $\jaco(\img)$ for the $i$th block of the input \\        
        $\jacos(\img)\in\mathbb{R}^{\numrma\times \numpix}$ & Sketched version of $\jaco(\img)$\\
        $\jacos^{(i)}(\img)\in \mathbb{R}^{n_s\times n_{pb}}$ & Columns of $\jacos(\img)$ for the $i$th block of the input \\
        
        \bottomrule
    \end{tabular}
    \caption{List of symbols and their meaning.}
    \label{tab:notations}
\end{table}

\section{Rate-distortion optimization} 
\label{sec:preliminaries}
Let $\img$ be an image with $n_p$ pixels and $\cimg(\thetavec)$ its compressed version using parameters $\thetavec\in\Theta$, where $\Theta\subset\mathbb{N}^{n_b}$ is the set of all possible operating points and $\nummb$ is the number of blocks. Assume every entry of $\thetavec$ takes values in the set $\lbrace 1, \hdots, n_r\rbrace$, where $n_r$ denotes the number of RDO options. Given blocks of size $n_{pb}$, $\img_i\in\mathbb{R}^{n_{pb}}$ for $i = 1, \hdots, \nummb$, we aim to find parameters $\thetavec^\star$ satisfying \cite{ortega_rate-distortion_1998}:
\begin{equation}
    \thetavec^\star = \argmin_{\thetavec \in \Theta} \, d( \cimg(\thetavec), \img) + \lambda \, \sum_{i = 1}^{n_b}\, r_i(\cimg_i(\thetavec)),
\end{equation}
where $d(\cdot, \cdot)$ is the distortion metric, $r_i(\cdot)$ is the rate for the $i$th coding unit, and $\lambda\geq 0$ is the Lagrange multiplier controlling the RD trade-off. We are especially interested in distortion metrics that decompose as the sum of block-wise distortions,
\begin{equation}
\label{eq:local_gen}
d(\cimg_1(\thetavec), \hdots, \cimg_{\nummb}(\thetavec), \img_1, \hdots, \img_{\nummb}) = \sum_{i = 1}^{\nummb} \, d_i(\cimg_i(\thetavec), \img_i), 
\end{equation}
which is true for SSE but may not hold for other metrics. When each coding unit can be optimized independently, we obtain $\cimg_i(\thetavec) = \cimg_i(\theta_i)$ \cite{ortega_rate-distortion_1998, fernandez2025rate}, which leads to 
\begin{equation}
\label{eq:final_form}
  \theta_i^\star = \argmin_{\theta_i \in \Theta_i} \, d_i(  \cimg_i(\theta_i), \img_i) + \lambda \, r_{i}(\cimg_i(\theta_i)),  
\end{equation}
for $i = 1, \hdots, \nummb,$ where $\Theta_i$ is the set of all  parameters for the $i$th block. This is the RDO formulation most video codecs solve \cite{ortega_rate-distortion_1998}. A practical way to choose $\lambda$ is \cite{sullivan_rate_1998}:
\begin{equation}
\label{eq:og_multiplier}
\lambda = c \, 2^{(\mathrm{QP}-12) / 3},
\end{equation}
where $\mathrm{QP}$ is the quality parameter, and $c$ varies with the type of frame and content \cite{ringis_disparity_2023}. This work aims to replicate this block-level RDO formulation in a CM scenario using distortion metrics derived from computer vision tasks.

\section{Feature-preserving RDO}
\label{sec:feature_rdo}
We formulate our CM problem in \autoref{sec:feat_extract}. Then, we introduce linearization (\autoref{sec:taylor}), block-wise approximation (\autoref{sec:blockwise_localization}), and \newtext{methods to sketch the} Jacobian (\autoref{sec:rma}).

\subsection{Problem formulation}
\label{sec:feat_extract} 
Let the \emph{feature extractor} be a function $f(\cdot)$ mapping images with $n_p$ pixels to $n_{\sf f}$-dimensional representations. In this work, we focus on feature extractors comprising a set of earlier layers of a DNN-based system. Assume we target $n_t$ tasks, each of them with task loss $\ell_k(\cdot, \cdot)$, e.g., cross-entropy loss (CEL) or SSE, for $k = 1, \hdots, n_t$. Given an input $\bt x$ with ground truth labels $\bt y_k$ and DNNs $g_k(\cdot)$, the loss is given by $\ell_k(\bt y_k, g_k(\bt x))$, for all $k$. 
Let the feature extractor be shared, such that the DNNs can be written as $g_k(\img) = h_k(f(\img))$ for all $k$\footnote{\newtext{This is only an assumption to facilitate the formulation. Experimentally, we show that optimizing lossy encoding based on the layers $f(\cdot)$ preserves task performance even if the target DNN does not include those layers, as long as it is trained to solve the same or a related task as the one using $f(\cdot)$.}}.  
We evaluate the degradation in performance due to compression via the consistency loss  $r_k(\cimg, \img) \doteq \norm{ \ell_k(\bt y_k, g_k(\cimg)) - \ell_k(\bt y_k, g_k(\img ))}_2^2$, for all $k$, which is the metric of choice when approximating the output of one system using another–e.g., network distillation  \cite{tarvainen2017mean}. Next, we relate consistency loss to feature distance (FD).

\begin{proposition}
\label{prop:feature-dist}
    Let $\ell_k(\cdot, \cdot)$ and $h_k(\cdot)$ be Lipschitz continuous functions with constants $L_k$ and $H_k$, respectively, for $k = 1, \hdots, n_t$. Then,
    \begin{equation}
    r_k(\cimg, \img) \leq H^2_k L^2_k \, \|f(\cimg)-f(\img)\|^2_2, \quad \text{for } k=1\hdots, n_t.
    \end{equation}    
\end{proposition}
\begin{proof}
    The result follows by consecutively applying Lipschitz continuity for $\ell_k(\bt y_k, \cdot)$ and $h_k(\cdot)$ for all $k$.
\end{proof}
The task losses (SSE and CEL) \cite{bethune2022pay} and neural networks \cite{virmaux2018lipschitz}  we consider in this work are Lipschitz continuous. From \autoref{prop:feature-dist}, we conclude that \emph{minimizing feature distance can preserve task performance}. Hence, we write the CM problem as minimizing the FD subject to a rate constraint:
\begin{equation}
    \thetavec^{\star} = \argmin_{\thetavec\in\Theta} \  \norm{ f(\cimg(\thetavec))-f(\img)}_2^2 + \lambda \, \sum_{i = 1}^{\nummb}\, r_i(\cimg_i(\theta_i)).
    \label{eq:RDO-with-feature-dist}
\end{equation}
\newtext{We call this formulation FD-RDO.} This approach allows choosing a feature extractor based on prior task knowledge. For instance, when pre-trained early layers are used across a series of tasks–e.g., transfer learning \cite{jain2023data}–we \newtext{construct a FD with these early layers to preserve performance across tasks.}


FD does not satisfy the locality property in \eqref{eq:local_gen}: evaluation requires the complete decoded image in the pixel domain. As a result, \newtext{FD-RDO} involves an iterative workflow of encoding, decoding, and feature distance evaluation, which is computationally impractical. Next, we propose an alternative solution.


\subsection{Linearizing the feature extractor}
\label{sec:taylor}
We assume the feature extractor has third-order partial derivatives almost everywhere, which is satisfied by the DNNs we consider in this work \cite{jacot_neural_2018}. Define the Jacobian matrix of $f(\cdot)$ evaluated at the input image $\img$ as $\jaco(\img)\in\mathbb{R}^{\numfeat \times \numpix}$, where:
\begin{equation}
    J_{ij}(\img) = \pdv{f_i(\img)}{x_j}, \quad i= 1, \hdots, \numfeat, \ \ j = 1, \hdots, \numpix.
\end{equation}
Rewriting $\cimg(\thetavec) = \img + (\cimg(\thetavec) - \img)$, we can apply Taylor's expansion to the feature extractor around $\img$ (\autoref{fig:taylor_expansion}):
\begin{equation}
\label{eq:taylor}
    f(\cimg(\thetavec)) = f(\img) + \jaco(\img) (\pixres) + o\big(\norm{\pixres}_2^2\big),
\end{equation}
where \newtext{$o(x)$ goes to zero at least as fast as $x$. \newtext{The entry of the Jacobian $J_{ij}(\img)$ is the contribution of the $j$th pixel to the $i$th feature. This expansion yields an expression for the features in terms of the Jacobian. Now, we use this expansion to construct an approximation for the FD in terms of the Jacobian.} In particular, we \newtext{approximate the FD $d(\cimg(\thetavec), \img) =  \norm{f(\cimg(\thetavec))-f(\img)}_2^2$ by a quadratic metric}\footnote{This approximation follows because $d(\img, \img) = 0$ and $d(\bt y, \img)$ has a minimum when $\bt y = \img$, therefore $\nabla_{\mathbf{y}}d(\bt y, \bt x)\vert_{\mathbf{y} = \mathbf{x}} = \bt 0$.}:
\begin{equation}
    d(\cimg(\thetavec), \img) = \frac{1}{2}(\cimg(\thetavec)-\img)^\top \bt H(\img)(\cimg(\thetavec)-\img) + o(\norm{\cimg(\thetavec)-\img}_2^3),
\end{equation}
where $\bt H(\img)$ is the Hessian matrix, which can be defined as
\begin{equation}
\label{eq:hessian}
    H_{ij}(\img) \doteq \frac{\partial^2 d(\bt y, \img)}{\partial y_i\partial y_j}\bigg\vert_{\bt y=\img}, \ i= 1, \hdots, \numfeat, \ j = 1, \hdots, \numpix.
\end{equation}
It can be proven that $\bt H(\img) = 2\jaco(\img)^\top\jaco(\img)$ \cite{do_carmo1992riemannian}. This expression is equivalent to the result we obtain from linearizing the feature extractor as in \eqref{eq:taylor} into the expression for the FD:
\begin{equation}
\label{eq:jaco}
    \norm{f(\cimg(\thetavec))-f(\img)}_2^2 \cong  \norm{\jaco(\img)(\pixres)}_2^2,
\end{equation}
where $\cong$ denotes high bit-rate convergence \cite{gish1968asymptotically}. Although this theoretical derivation relies on a high bit-rate assumption, we will test and verify experimentally our approximations in operational regimes, which include low and medium bit-rate scenarios, in \autoref{fig:rd_fd_idse_avg} and \autoref{sec:diff_codecs}. The cubic error bound can be justified from Taylor's expansion:
\begin{equation}
    \big\vert  d(\cimg(\thetavec), \img) - \norm{\jaco(\img)(\cimg(\thetavec)-\img)}_2^2 \big\vert \leq \frac{M}{6} \norm{\cimg(\thetavec) - \cimg}_2^3,
\end{equation}
with $M$ bounding the third-order derivative of the distortion metric $d^{(3)}(\tilde{\img}(w), \img)$ for all $\tilde{\img}(w) = w\img + (1-w)\cimg$ with $w\in[0, 1]$ \cite{do_carmo1992riemannian}}. As a result of this approximation, the RDO  of \eqref{eq:RDO-with-feature-dist} becomes
\begin{equation}
\label{eq:optimization_jaco}
    \thetavec^{\star} = \argmin_{\thetavec\in\Theta} \,\norm{\jaco(\img)(\pixres)}_2^2 + \lambda \, \sum_{i = 1}^{\nummb}\, r_i(\cimg_i(\theta_i)).
\end{equation}
Since this formulation requires the whole image, 
block-level bit allocation is still impractical. 
The next section addresses this problem by localizing the metric.
\begin{figure}[t]
    \centering
    \includegraphics[width=\linewidth]{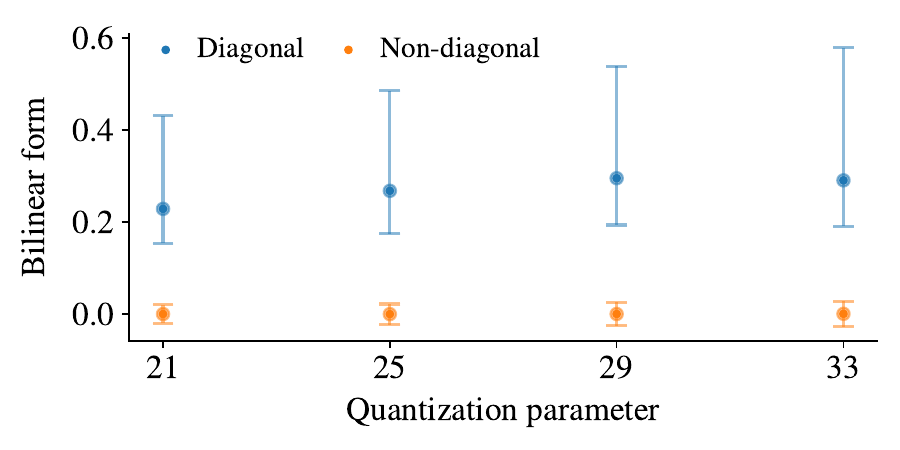}
    \caption{Median value with $15$th and $85$th percentiles of the diagonal/off-diagonal terms of the bilinear forms in \eqref{eq:block_loss}, after normalization, computed with $1000$ blocks of size $16\times 16$ from the COCO dataset using SSE-RDO AVC.}
    \label{fig:blockwise_assump}
\end{figure}
\begin{figure*}[t]
    \centering
\includegraphics[width=\linewidth]{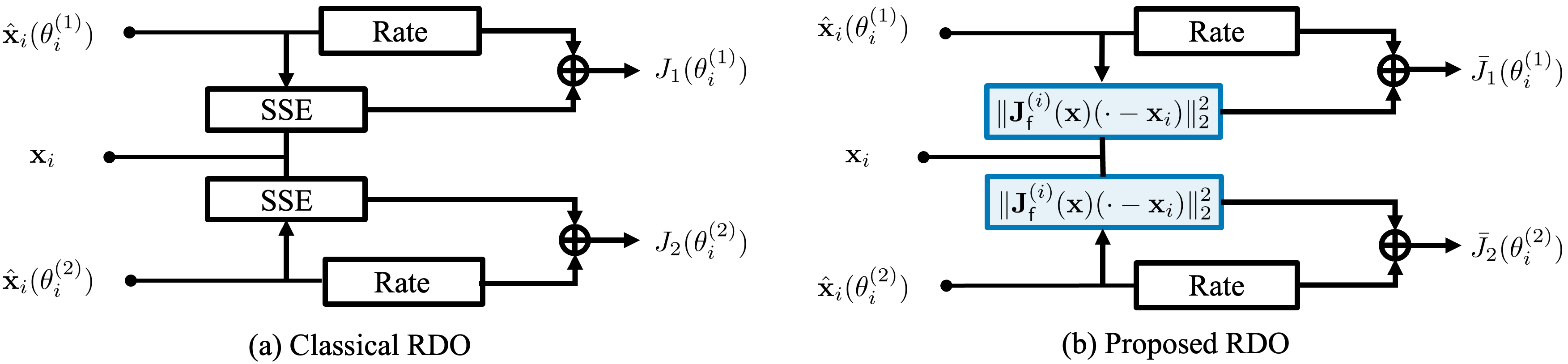}
    \caption{Process of comparing two RDO options, $\theta^{(1)}_i$ and $\theta^{(2)}_i$, using (a) classical SSE-RDO and (b) our proposed IDSE-RDO.}
\label{fig:rdo_blocks}
\end{figure*}
\subsection{Block-wise localization}
\label{sec:blockwise_localization}
We first write the Jacobian in terms of the sub-matrices corresponding to the pixels in each block:
\begin{equation}
\label{eq:blockwise}
    \bt J_{\mathsf{f}}(\img) = \begin{bmatrix} \jaco^{(1)}(\img) & \jaco^{(2)}(\img) & \hdots & \jaco^{(n_b)}(\img)\end{bmatrix}, 
\end{equation}
with $\jaco^{(i)}(\img)\in\mathbb{R}^{\numfeat\times n_{pb}}$, for $i = 1, \hdots, n_b$. \newtext{Define the overall and block-level quantization errors as  $\bt e(\thetavec)=\pixres$ and $\bt e_i(\theta_i) \doteq \cimg_i(\theta_i) - \img_i$, for all $i$. 
We can write the loss as a sum of bilinear forms:
\begin{equation}
\label{eq:block_loss}
\norm{\jaco(\img)\bt e(\thetavec)}_2^2 =  \hspace{0.75em} \sum_{j = 1}^{n_b}\, \sum_{i = 1}^{n_b} \, \bt e_i(\theta_i)^\top\jaco^{(i)}(\img)^\top\jaco^{(j)}(\img)\bt e_j(\theta_j).
\end{equation}
Under} the high bit-rate model \cite{gish1968asymptotically}, quantization errors across blocks are uncorrelated, and the expectation of cross-block terms is zero, 
\begin{equation}
\label{eq:block_loss_uncorrelated}
\newtext{\mathbb{E}_{\footnotesize \bt e_i, \bt e_j}( \bt e_i^\top\jaco^{(i)}(\img)^\top\jaco^{(j)}(\img)\bt e_j ) = 0, \quad \text{if } i\not = j.}
\end{equation}
This suggests that the off-diagonal terms contribute minimally compared to the diagonal terms. As an empirical test, we compute $\bt b_i(\img) = \jaco^{(i)}(\img)\bt e_i(\theta_i)/(\norm{\bt e_i(\theta_i)}_2\norm{\jaco(\img)}_F)$ using residuals from multiple quantization levels. Then, we obtain the diagonal $\bt b_i(\img)^\top \bt b_i(\img)$ and off-diagonal $\bt b_j(\img)^\top \bt b_i(\img)$ bilinear forms. The diagonal terms dominate the cross-terms (\autoref{fig:blockwise_assump}). \newtext{We propose using the aforementioned block approximation to  the linearized metric}
\begin{equation}
    \label{eq:localization}
    \norm{\jaco(\img)(\pixres)}_2^2 \approx \sum_{i = 1}^{\nummb} \,\|\jaco^{(i)}(\img)(\pixresi)\|_2^2.
\end{equation}
We call the right-hand side of \eqref{eq:localization} \emph{input-dependent squared error} (IDSE). 
Since IDSE is a sum of block-wise distortions,  we can convert \eqref{eq:optimization_jaco} into  the  RDO problem: 
\begin{equation}
\label{eq:optimization_jaco_block}
    \thetavec^{\star} = \argmin_{\thetavec\in\Theta} \,\sum_{i = 1}^{\nummb} \,\|\jaco^{(i)}(\img)(\pixresi)\|_2^2+ \lambda \, \sum_{i = 1}^{\nummb}\, r_i(\cimg_i(\theta_i)),
\end{equation}
\newtext{which has the same form as  \eqref{eq:final_form} and can be  solved per block:}
\begin{equation}
\label{eq:bw_rdo}
    \theta^{\star}_i = \argmin_{\theta_i\in\Theta_i} \, \|\jaco^{(i)}(\img)(\pixresi)\|_2^2 + \lambda \, r_i(\cimg_i(\theta_i)),
\end{equation}
for $i = 1, \hdots, \nummb$.
\autoref{fig:rdo_blocks} compares this RDO formulation, \newtext{which we term IDSE-RDO}, to SSE-RDO. \newtext{The transition from the linearized FD \eqref{eq:optimization_jaco} into its block-wise approximation \eqref{eq:optimization_jaco_block} can be viewed mathematically as approximating the quadratic form $\bt e(\thetavec)^{\top} \bt H(\img) \bt e(\thetavec)$ (see \eqref{eq:hessian}) by $\bt e(\thetavec)^{\top} {\bt P}(\img) \bt e(\thetavec)$, where ${\bt P}(\img)$ is a  block-diagonal matrix with block diagonal terms $\jaco^{(i)}(\img)^\top\jaco^{(j)}(\img)$.} 
\newtext{Besides \eqref{eq:block_loss_uncorrelated}, we expect this approximation to be also valid if the features are approximately localized, i.e., perturbations in an image block affect only a small contiguous set of entries in the feature vector. This condition would be true for shallower networks such as convolutional layers \cite{lecun2002gradient}.}

\subsection{Randomized IDSE approximation}
\label{sec:rma}
\newtext{The approximations of the previous section allow us to re-formulate FD-RDO in a way that can be used in block-based video codecs in CM applications. However, solving \eqref{eq:bw_rdo}  is not feasible under the computational constraints of real-world coding applications. The Jacobian in \eqref{eq:blockwise} is a large $\numfeat \times \numpix$ matrix.  Thus, in order to compute it we need $\numfeat$ backward passes (i.e., one gradient with respect to the input for each entry in the feature vector at the output of the DNN), \newtext{with $\numfeat$ typically a very large number\footnote{For the features considered in this work, $\numfeat$ ranges from $10^5$ to $10^8$.}}. 
Even if we manage to compute (and can store) the complete Jacobian, evaluating the distortion in \eqref{eq:bw_rdo} requires $\numfeat$ inner products (for the matrix-vector product with $\jaco^{(i)}(\img)$) for each of the $n_r$ possible RDO candidates, and this operation has to be repeated for every block. 
 In contrast, to evaluate the distortion in SSE-RDO, we only require computing the norm of an $n_{pb}$-dimensional vector for each of the $n_r$ RDO choices. }

\newtext{
We propose using   \emph{randomized numerical linear algebra methods} to approximately solve the RDO problem in \eqref{eq:bw_rdo}, without explicitly computing or storing the full Jacobian, nor performing operations of dimension $\numfeat$ to evaluate the distortion metric. 
We leverage random sketches, where vectors are multiplied by a random matrix that approximately preserves their norms \cite{martinsson2020randomized}, allowing us to approximate the distortions in \eqref{eq:bw_rdo} with a fraction of the computational cost.  
We compute sketched Jacobians  $\jacos^{(i)}(\img)\doteq \bt S\jaco^{(i)}(\img)$, for $i = 1, \hdots, n_b$, where $\bt S \in\mathbb{R}^{n_{s}\times \numfeat}$ is a sketching matrix such that $n_{s} \ll \numfeat$.  
By choosing   $\bt S$ based on the Johnson–Lindenstrauss lemma \cite{johnson1984extensions, achlioptas_database_2003}, each norm  $\|\jaco^{(i)}(\img)(\pixresi)\|_2^2$ can be well approximated by $\|\jacos^{(i)}(\img)(\pixresi)\|_2^2$ in a  computationally efficient way.  Then, we can approximate \eqref{eq:bw_rdo} by solving
\begin{equation}
\label{eq:idse-rdo}
    \theta_i^\star = \argmin_{\theta_i\in\Theta_i} \|\jacos^{(i)}(\img)\l \pixresi\r\|_2^2 + \lambda \, r_i(\cimg_i(\theta_i)).
\end{equation}
As a result, we can store only the sketched Jacobian, which is cheaper than storing the full Jacobian matrix. Sketching can be done just once, right before starting the encoding or RDO processes, and it can be done efficiently via autodiff \cite{paszke_automatic_2017} on the feature extractor. Next, we justify the approximation in \eqref{eq:idse-rdo} and detail how to efficiently obtain the sketched Jacobians.}

\subsubsection{Jacobian sketching algorithm}
\newtext{Since we use the same sketching matrix for all $\jaco^{(i)}(\img)$, we can rely on \eqref{eq:blockwise} to write 
\begin{align}
  &\jacos(\img)=\bt S\jaco(\img) =   \bt S \begin{bmatrix} \jaco^{(1)}(\img) & \hdots & \jaco^{(n_b)}(\img)\end{bmatrix} =  \nonumber \\
  &\Big[\bt S\jaco^{(1)}(\img) \ \hdots \ \bt S\jaco^{(n_b)}(\img)\Big] = \Big[ \jacos^{(1)}(\img) \ \hdots \ \jacos^{(n_b)}(\img)\Big]. 
\end{align}
Thus, the sketched Jacobians for each block $\jacos^{(i)}(\img)$ can be computed simultaneously from a sketch of the whole-image Jacobian, $ \jacos(\img)$. The sketched Jacobian for a given $\bt S$  is found (see \autoref{alg:jacobian}) by  
obtaining the features $f(\img)$ for a given input $\img$, evaluating $\bt q(\img) = \bt Sf(\img)$, and computing, using autodiff \cite{paszke_automatic_2017}, the gradient of each of the entries of the reduced vector $\bt q(\img)$ with respect to $\img$.}
\newtext{Thanks to sketching}, we can reduce the number of computations by taking derivatives from a lower-dimensional vector $\bt q(\img)$,  \newtext{since the gradient of the $i$th entry of $\bt q(\img)$ with respect to $\img$ satisfies:
\begin{equation}
    \nabla_{\mathbf{x}} q_i(\img) = \nabla_{\mathbf{x}}\, \bt s_i^\top f(\img) = \sum_{j = 1}^{\numfeat} s_{ij}\nabla_{\mathbf{x}}f_j(\img) = \bt s_i^\top \jaco(\img),
\end{equation}
where $\bt s_i$ is the $i$th row of $\bt S$ and we assume $\nabla_{\mathbf{x}}f_j(\img)$ is a row vector, for $j = 1, \hdots, \numfeat$. By row-stacking the result of computing the gradient for each entry, we obtain $\begin{bmatrix}\nabla_{\mathbf{x}} q_1(\img)^\top & \hdots & \nabla_{\mathbf{x}} q_{\numrma}(\img)^\top \end{bmatrix}^\top = \bt S\jaco(\img) = \jacos(\img)$. \newtext{Hence, $\jacos(\img)$} follows from $\numrma$ backward passes by computing derivatives in a $\numrma$-dimensional space instead of computing the full Jacobian, which requires $\numfeat$ passes in a $\numfeat$-dimensional space.}
\subsubsection{Selecting $n_s$ to approximate  \eqref{eq:bw_rdo} with \eqref{eq:idse-rdo}}
 \newtext{To solve \eqref{eq:bw_rdo}, we need to compute the RD cost for each $\theta_i \in \Theta_i$. Since the set of RDO candidates  $\Theta_i$ has cardinality $n_r$, we need to be able to approximate $\|\jaco^{(i)}(\img)(\pixresi)\|_2^2$ precisely enough to distinguish between these choices. We propose an approach based on the Johnson–Lindenstrauss (JL) lemma \cite{johnson1984extensions, achlioptas_database_2003}.
 \begin{lemma}[Johnson-Lindenstrauss]
 \label{lemma:JL}
     Let $\mathcal{X} \subset \mathbb{R}^d$ be a finite set with cardinality $p = \vert \mathcal{X} \vert$. For any $\epsilon \in (0,1)$, there exists a matrix $\bt S \in \mathbb{R}^{m \times d}$, such that
     \begin{equation}
         (1-\epsilon)\Vert \bt y - \bt z\Vert^2 \leq \Vert \bt S(\bt y -\bt z)\Vert^2\leq (1+\epsilon)\Vert \bt y - \bt z\Vert^2, 
     \end{equation}
    for all $\bt y, \bt z \in \mathcal{X}$,  provided that
     \begin{equation}
         m \geq c\log(p)/\epsilon^2,
     \end{equation}
     where $c$ is a fixed numerical constant.
 \end{lemma}
 We consider the matrix construction given by $\bt S = (1/\sqrt{m}) \bt R$ \cite{achlioptas_database_2003}, where $\bt R$ is an $m \times d$ random matrix, with entries in $\lbrace 1, -1 \rbrace$ with equal probability. This construction satisfies the JL lemma with probability at least $1- 2\exp(-m \eta(\epsilon))$, where $\eta(\epsilon)$ depends on the distribution of $\bt S$\cite{jacques2015quantized}. This choice balances performance and memory efficiency \cite{fernandez2024feature}.}

\newtext{To  solve \eqref{eq:bw_rdo} we would need to evaluate the distortion  $\|\jaco^{(i)}(\img)(\pixresi)\|_2^2 $ for each $\theta_i \in \Theta_i$. Instead, we approximate those distortions using the JL lemma and solve \eqref{eq:idse-rdo}.  Define the set of  $n_r +1$ points $\mathcal{X}_i=\lbrace \jaco^{(i)}(\img) \hat{\bt x}_i(\theta_i): \theta_i \in \Theta_i \rbrace \cup \lbrace \jaco^{(i)}(\img) \bt x_i \rbrace $. Then, by applying the JL lemma to $\mathcal{X}_i$ for a given  $\epsilon \in (0,1)$, we choose a sketching dimension  
\begin{equation}
\label{eq:samples}
\newtext{n_{s} \geq c\log(n_r+1)/{\epsilon}^2},
\end{equation}
which depends on the number of RDO candidates for each block, but is independent of the block size and the number of features. 
With this $n_s$, given a random matrix $\bt S\in\mathbb{R}^{\numrma\times \numfeat}$  with entries taken from the set $\lbrace 1/\sqrt{n_s}, -1/\sqrt{n_s} \rbrace$, we have 
\begin{multline}
\label{eq:jonlind}
    (1-\epsilon)\|\jaco^{(i)}(\img)(\pixresi)\|_2^2 \leq \|\jacos^{(i)}(\img)(\pixresi)\|_2^2  \\ \leq (1+\epsilon)\|\jaco^{(i)}(\img)(\pixresi)\|_2^2,
\end{multline}
with probability at least $1- 2\exp(-n_{s} \eta(\epsilon))$, allowing us to use the sketched distortions for IDSE-RDO. 
Note that the condition on $n_s$ \eqref{eq:samples} and the approximation tolerance ($\epsilon$) are the same for all blocks.}
\subsubsection{Discussion: complexity vs approximation}
\newtext{In our approach, increasing $\numrma$ improves the quality of the sketch by allowing for a smaller tolerance $\epsilon$ in \eqref{eq:jonlind}. Yet, $n_s$ also controls the computational complexity of our method. In particular,  1) the complexity of obtaining $\jacos(\img)$ increases linearly with $n_s$, since we need a backward pass for each element in the $\numrma$-dimensional space we obtain after sketching, and 2) the computational complexity of evaluating the distortion metric for IDSE-RDO in \eqref{eq:idse-rdo} increases linearly with $\numrma$ because we need as many inner products with the error as rows in $\jacos^{(i)}(\img)$. Similarly, we need to store $\jacos(\img)$, which requires $\numrma\times\numpix$  memory entries. Since the final goal is not to estimate the Jacobian—or even the IDSE—but rather to rank RDO candidates correctly while accounting for the downstream task, a small $n_s$ is often sufficient, as we show experimentally in \autoref{sec:exper} (cf.~\autoref{tab:num_samples}).}

\begin{figure}
    \centering
    \includegraphics[width=\linewidth]{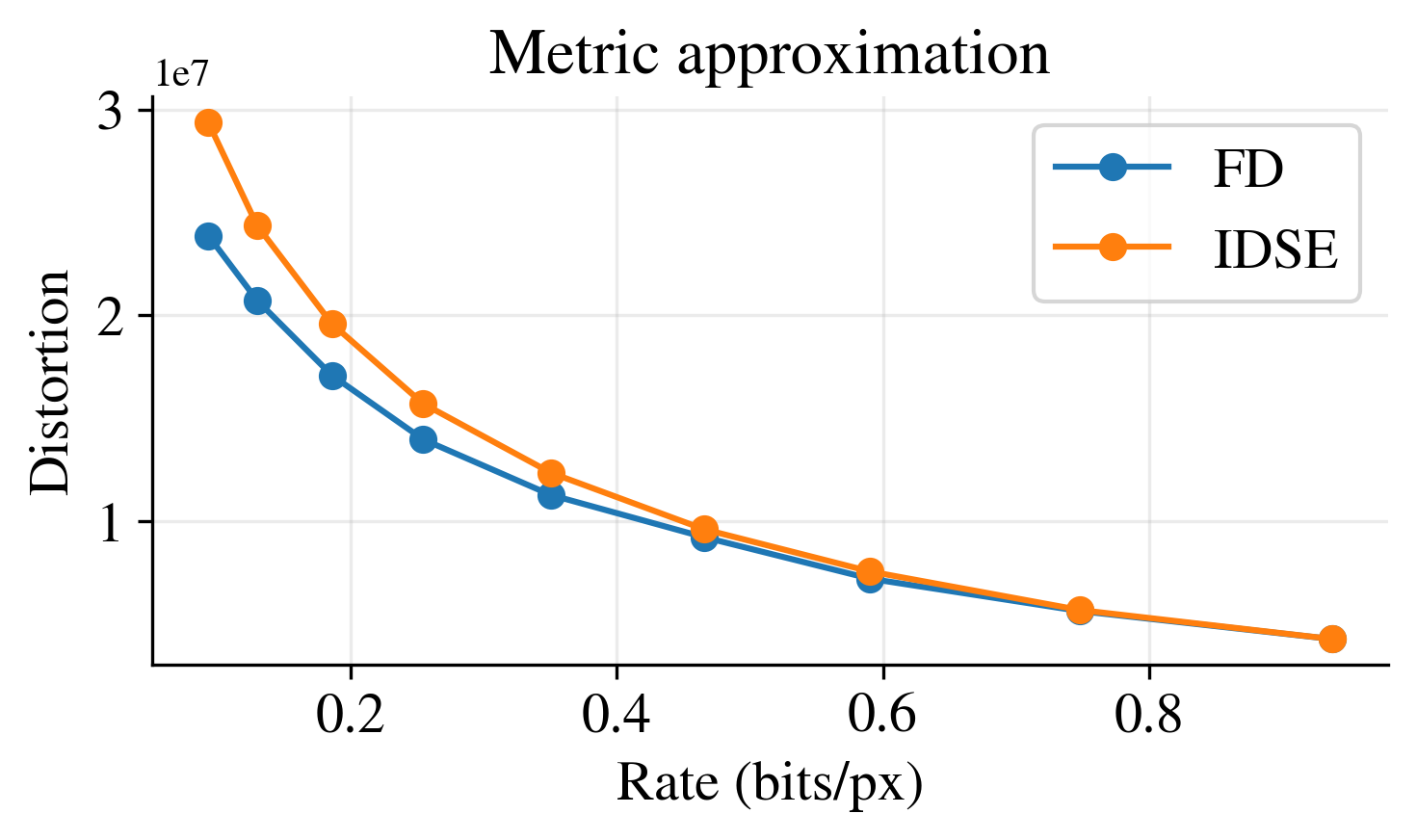}
    \caption{\newtext{Average RD curves for $1000$ images from the COCO dataset \cite{lin_microsoft_2014} compressed using AVC with QP ranging between $31$ and $47$. IDSE converges to FD as the rate increases.}}
    \label{fig:rd_fd_idse_avg}
\end{figure}

\newtext{To assess empirically the quality of all our approximations, we depict the average RD curves for a set of images compressed using AVC in \autoref{fig:rd_fd_idse_avg}. As a distortion metric, we consider both the sketched IDSE evaluated at the frame-level with $\numrma=8$ and FD. 
We observe that the sketched IDSE converges to the FD as the rate increases, which validates our approximation based on the Taylor expansion in \eqref{eq:taylor} and the block-diagonal approximation in \eqref{eq:localization}. 
We emphasize that the goal of our approach 
is not to minimize the approximation error between the true FD and the IDSE, but rather to provide a metric that behaves like FD but is easier to compute and incorporate into RDO in a block-based codec. As long as the approximation preserves RD ordering, the system performance remains robust even if the approximation error is non-negligible in absolute terms.} Next, we discuss how to incorporate IDSE into a coding pipeline to balance visual quality with CV performance.

\section{Image coding with IDSE}
\label{sec:integration}
We discuss SSE regularization (\autoref{sec:reg}), transform domain IDSE (\autoref{sec:transform}), and IDSE-RDO with Lagrangian selection (\autoref{sec:lag_sel}). \autoref{sec:complexity} summarizes complexity.

 \subsection{SSE regularization}
 \label{sec:reg}
Our proposed loss can be combined with SSE to balance visual quality and CV performance.
For a given frame, start with the RDO problem in \eqref{eq:optimization_jaco} and let $\mathrm{SSE}_{\mathrm{max}}$ be the maximum admissible SSE. 
We want to solve:
\begin{multline}
    \thetavec^\star = \argmin_{\thetavec\in\Theta} \,\|\jaco(\img)\l \pixres\r\|_2^2 + \lambda \, \sum_{i = 1}^{n_b}\, r_i(\cimg_i(\theta_i)), \\
    \text{such that} \ \norm{\pixres}_2^2 \leq \mathrm{SSE}_{\mathrm{max}}.
\end{multline}
Applying Lagrangian relaxation \cite{everett_generalized_1963} and grouping together all the terms other than the rate:
\begin{equation}
\label{eq:distortion_balance}
    d(\cimg(\thetavec), \img) = \|\jaco(\img)\l \pixres\r\|_2^2 + \tau \, \norm{\pixres}_2^2,
\end{equation}
where $\tau\geq 0$ is the regularization parameter. Expanding \eqref{eq:distortion_balance},
\begin{equation}
\label{eq:balance}
d(\cimg(\thetavec), \img) = (\cimg(\thetavec)-\img)^\top \bt Q_{\tau}(\img) (\cimg(\thetavec)-\img),
\end{equation}
with $\bt Q_{\tau}(\img) = \jaco(\img)^\top\jaco(\img) + \tau \, \bt I$, for $i = 1, \hdots, n_b$, which can be interpreted as a Tikhonov regularization, where larger $\tau$ gives more importance to SSE.    

Additionally, since the magnitude of the Jacobian changes with the feature extractor, 
we will need to choose different values for $\tau$ in \eqref{eq:distortion_balance} for each $f(\cdot)$ and $\img$, even if the relative importance of task accuracy and image representation is fixed.  
To simplify the selection of $\tau$, making it more consistent and interpretable across different $f(\cdot)$ and $\img$, 
we define $\tau = \tilde{\tau}\alpha$, where $\tilde{\tau}$ is a normalization factor to make the contributions of IDSE and SSE in \eqref{eq:distortion_balance} equal. 
As a heuristic, for the largest desirable quantization step size, $\Delta_{\max}$, for a given application (typically smaller than the maximum step size allowed by the codec), 
we choose $\tilde{\tau}$ so that, for $\Delta_{\max}$, $\| \jaco(\img)(\cimg(\thetavec) - \img)\|_2^2 = \tilde{\tau}\norm{\pixres}_2^2$, giving both terms in \eqref{eq:distortion_balance} equal contribution to the final loss when $\alpha=1$ for any $f(\cdot)$ and $\img$.
\begin{proposition}
    For a given matrix $\jaco(\img)$,
    \begin{equation}
    \label{eq:init_tau}
        \tilde{\tau} = \norm{\jaco(\img)}^2_2,
    \end{equation}
    ensures $\max_{\thetavec}\| \jaco(\img)\l \pixres\r\|_2^2 = \tilde{\tau} \max_{\thetavec}\norm{\pixres}_2^2$.
\end{proposition}
\begin{proof}
    The regularizer is given by the ratio:
    \begin{equation}
        \tilde{\tau} = \max_{\thetavec}\| \jaco(\img)\l \pixres\r\|_2^2 / \max_{\thetavec}\norm{\pixres}_2^2.
    \end{equation}
    By definition of spectral norm, and assuming the set of coding options is dense, there is an error that reaches the upper bound:
    \begin{equation}
    \| \jaco(\img)\l \pixres\r\|_2^2 \leq \| \jaco(\img)\|_2^2\norm{\pixres}_2^2.
    \end{equation}
    The result follows since $\jaco(\img)$ does not depend on $\thetavec$.
\end{proof}
\newtext{In practice, we use $\norm{\jacos(\img)}_2^2$ as an estimator for $\norm{\jaco(\img)}_2^2$.} This choice of $\tau$ can adapt to the characteristics of the content and the feature extractor. In our experiments, we test different values of $\alpha$ in $\tau = \alpha\tilde{\tau}$ to explore the trade-off between visual quality and downstream task performance (cf.~\autoref{fig:sweep_frext}).

\subsection{Transform domain evaluation}
\label{sec:transform}
    Let $\bt U$ be an orthogonal transform, such as the discrete cosine transform (DCT) \cite{strang_discrete_1999} or the asymmetric discrete sine transform (ADST) \cite{fernandez2024fast}. Define $\bt y_i \doteq \bt U^\top \bt x_i$, with $\hat{\bt y}_i(\theta_i)$ its quantized version. Then, the regularized IDSE becomes:
\begin{equation}
    d(\hat{\bt y}_i(\theta_i), \bt y_i) = \|\bt B^{(i)}(\img) (\hat{\bt y}_i (\theta_i)-\bt y_i)\|_2^2 + \tau\norm{\hat{\bt y}_i (\theta_i)-\bt y_i}_2^2, 
\end{equation}
where $\bt B^{(i)}(\img) \doteq \jacos^{(i)}(\img)\bt U$, for $i = 1, \hdots, \nummb$, \newtext{and the SSE term follows from Parseval's identity.} Thus, RDO can be conducted directly in the transform domain:
\begin{equation}
\label{eq:transform_domain}
\theta_i^\star = \argmin_{\theta_i\in\Theta_i} \, d(\hat{\bt y}_i(\theta_i), \bt y_i)+ \lambda \, r_i(\hat{\bt y}_i(\theta_i)), 
\end{equation}
and the RD cost can be evaluated using \eqref{eq:transform_domain}, which \newtext{is} more efficient than using \eqref{eq:idse-rdo}. 

\subsection{RDO with IDSE}
\label{sec:lag_sel}
First, we discuss the choice of Lagrange multiplier. Then, we summarize the process of doing RDO with IDSE. The Lagrangian in \eqref{eq:og_multiplier} is based on a simplified logarithmic model relating the expected rate to the expected SSE. Since we are not using SSE directly, this relationship no longer holds. Instead, we can derive an expression for $\lambda$ based on the expected value of the regularized IDSE in a given block: at high rates \cite{gish1968asymptotically}, using a quantization step $\Delta$,
\begin{equation}
    D_i = \mathbb{E}_{\footnotesize \cimg, \img} (d(\cimg_i, \img_i)) =  \big(\mathbb{E}_{\footnotesize \img}\big(\| \jacos^{(i)}(\img)\|_F^2\big) + \newtext{n_{pb}}\tau\big) \frac{\Delta^2}{12},
\end{equation}
for $i = 1, \hdots, \nummb$, \newtext{where $n_{pb}$ denotes the number of pixels in a block.}
Assuming the block Jacobians in the image are i.i.d., we obtain $D = D_i$. The logarithmic model in \cite{sullivan_rate_1998}  states that $R(D) = a\newtext{n_{pb}}\log(b\newtext{n_{pb}} / D)$, where $a$ and $b$ are related to the entropy power of the source. Then,  $\lambda = D/(na)$, and using the relationship between $\mathrm{QP}$, $\Delta$, and $\lambda$ in AVC \cite{wiegand_overview_2003},
\begin{equation}
\label{eq:lagrangian}
    \lambda = c \, \l \sum_{i = 1}^{n_b} \, \|\jacos^{(i)}(\img)\|_F^2/(\newtext{n_{pb}}n_b) + \tau\r\, 2^{(\mathrm{QP}-12)/3}.
\end{equation}
Experimentally, setting $c$ as in \eqref{eq:og_multiplier} leads to good rate control.

\begin{algorithm}[t]
\renewcommand{\Require}[1]{\State \textbf{Input:} #1}
    \caption{Jacobian sketching}
    \label{alg:jacobian}
    \begin{algorithmic}[1]
        \Require Feature extractor $f(\cdot)$, num.~samples $\numrma$, image $\img$        
        \State Evaluate $f(\img)\in\mathbb{R}^{\numfeat}$ \Comment{Compute features}
        \State Draw $\bt S\sim\mathrm{Rademacher}(\numrma\times\numfeat)$ \Comment{$\numrma\times\numfeat$-matrix  $\pm1$}
        \State Compute $\bt q(\img) = \bt Sf(\img)$ \Comment{$\bt q(\img)\in\mathbb{R}^{\numrma}$}
        \For{$i = 1$ to $\numrma$}
            \State Get $\nabla_{\mathbf{x}} q_i(\img)$ via autodiff
        \EndFor
        \State $\jacos(\img) \gets \begin{bmatrix}\nabla_{\mathbf{x}} q_1(\img)^\top & \hdots & \nabla_{\mathbf{x}} q_{\numrma}(\img)^\top \end{bmatrix}^\top / \sqrt{n_s}$ 
        \State \Return $\jacos(\img)$
    \end{algorithmic}
\end{algorithm}

 \begin{figure}[t]
\centering
\includegraphics[width=\linewidth]{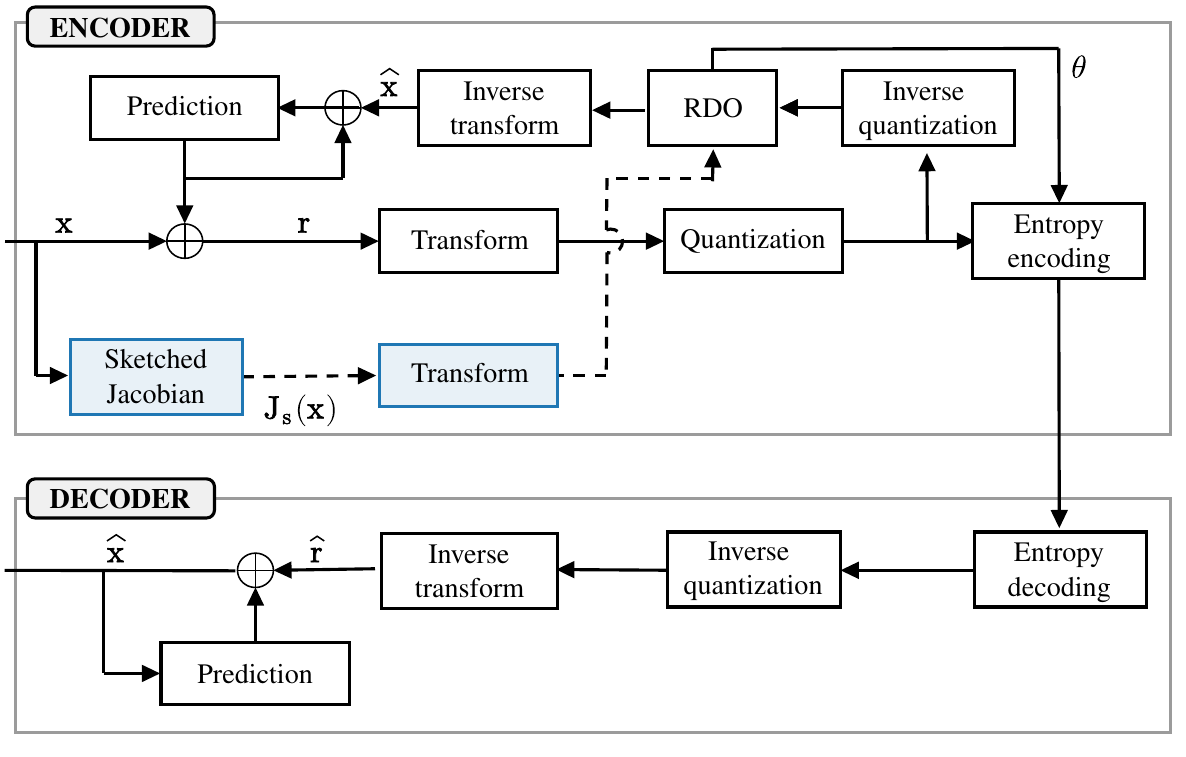}
\caption{Block diagram of the proposed codec, with the steps needed for IDSE-RDO in \textcolor{customblue}{\textbf{blue}.} Since we do not modify the decoder, it remains compatible with standardized codecs.}
\label{fig:bd}
\end{figure}

 \begin{figure*}[th]
     \centering
    \includegraphics[width=\linewidth]{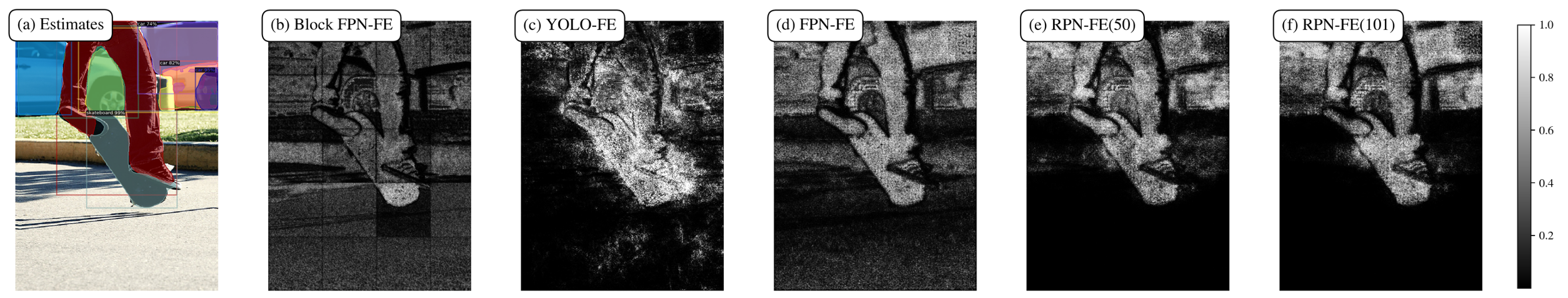}     \caption{(a) Mask R-CNN estimates; (b-f) $\mathrm{diag}(\jacos(\img)^\top \jacos(\img))$, reshaped and scaled, obtained by (b) localizing block-wise first and then expanding the metric, using blocks of size $128\times 128$ and a \fpnfe{}; (c-f) using our approach with different feature extractors. Lighter regions are more important; using the whole image and deeper models emphasizes relevant regions.}
     \label{fig:semantic}
 \end{figure*}

In our proposed codec (\autoref{fig:bd}), given an input $\img$, we compute $f(\img)$, obtain the sketched Jacobian $ \jacos(\img)$ (\autoref{alg:jacobian}), divide it into blocks $\jacos^{(i)}(\img)$, for $i = 1, \hdots, \nummb$, and compute its \newtext{modified version}, $\bt B^{(i)}(\img)$. Next, we encode each block of the input using different coding tools (\autoref{fig:rdo_blocks}) and perform RDO using \eqref{eq:transform_domain}. \newtext{This process is summarized in \autoref{alg:idse_rdo}}.

\begin{algorithm}[t]
\renewcommand{\Require}[1]{\State \textbf{Input:} #1}
\renewcommand{\Return}[1]{\State \textbf{Output:} #1}
\caption{\newtext{Compression with IDSE-RDO}}
\label{alg:idse_rdo}
\begin{algorithmic}[1]
    \Require Image $\img$, sketching dimension $n_s$, feature extractor $f(\cdot)$, QP, regularization parameter $\alpha$
    \State Compute $\jacos(\img)$ using \autoref{alg:jacobian}
    \State Split Jacobian to obtain $\jacos^{(i)}(\img)$, for $i = 1, \hdots, n_b$
    \State Compute $\tilde{\tau}$ as in \eqref{eq:init_tau} and $\lambda$ as in \eqref{eq:lagrangian}.
    \For{each block $\img_i$ in $\img$}
    \For{each RDO candidate $\theta_i \in \Theta_i$}
        \State $r_{i}(\cimg_i(\theta_i)) \gets \text{bitrate}(\cimg_i(\theta_i))$.    
        \State Compute error $\bt e_i(\theta_i) = \cimg_i(\theta_i) - \img_i$.        
        \State $d_{i}(\cimg_i(\theta_i), \img_i) \gets \|\jacos^{(i)}(\img)\bt e_i(\theta_i)\|_2^2 + \alpha\tilde{\tau}\|\bt e_i(\theta_i)\|_2^2$.
        \State $J_i(\theta_i) \gets d_{i}(\cimg_i(\theta_i), \img_i) + \lambda \, r_{i}(\cimg_i(\theta_i))$
    \EndFor
    \State $\theta_i^* \gets \argmin_{\theta_i} J_i(\theta_i)$
    \EndFor
    \State  Stack $\thetavec^* = [\theta_1^* \ \theta_2^* \, \hdots \, \theta_{n_b}^*]$
    \Return $\cimg(\thetavec^*)$
\end{algorithmic}
\end{algorithm}

\subsection{Computational complexity} 
\label{sec:complexity}
We compare the complexity of IDSE and block FD in terms of floating point operations (FLOPs), with runtimes given in \autoref{sec:runtime}. \newtext{We divide the complexity of our method into two parts: (i) computing the sketched Jacobian using the DNN, which, for a given image, has to be done only once before encoding, and (ii) given the sketched Jacobian, evaluating the metric for each RDO candidate. All the DNNs we consider resize the input images to a predefined fixed size before evaluating the neural network, both for training and for testing. Nonetheless, we ignore the cost of resizing because it is insignificant compared to the cost of evaluating the DNNs.} 

\newtext{Computing the sketched Jacobian requires a DNN forward pass and $\numrma$ backward passes–a backward pass having roughly twice the cost of a forward pass \cite{sepehri_hierarchical_2024}. 
Since images are first resized, the complexity of sketching the Jacobian depends on the size of the images after resizing. During encoding, once the Jacobian is available, evaluating the IDSE term in \eqref{eq:idse-rdo} involves computing the inner product of each of the $\numrma$ rows of the sketched Jacobian $\jacos(\img)$ with the error, which requires $\numrma  n_{pb}$ products, and then squaring and aggregating the results, which requires $\numrma$ products. These operations have to be repeated for each block and for each RDO candidate.} 

\newtext{Regarding block-FD\cite{fischer_video_2020}, block-wise coding decisions require DNN evaluation for individual blocks.} \newtext{We assume no resizing to be coherent with \cite{fischer_video_2020}, but even if resizing was performed, block FD would require block-wise DNN evaluations, so the DNN processing complexity in block FD scales with the size of the input image.} To compute block FD during encoding, we have to evaluate the DNN and compute the distance in feature space for each RDO candidate \newtext{and each block}.

\newtext{For a numerical comparison, }assume the input \newtext{before resizing} has $h\times w$ pixels, and after resizing to compute the Jacobian, we get images of  $h'\times w'$ pixels; also, let $n_r$ be the number of RDO candidates. Let $C$ be the cost of the forward pass in terms of floating point operations per pixel (FLOPs/px). We use the same feature extractor for both approaches. Using block FD, we require $h\times w\times (n_r+1)\times C$ FLOPs to evaluate the cost throughout the image. We require  $h'\times w'\times (2\numrma + 1)\times C$ FLOPs to sample the Jacobian. Assuming image sizes of $768\times 768$ pixels, resized images of size $224\times 224$, $n_r = 18$ ($9$ quantization steps and $2$ block partitions), and letting $\numrma = 4$, our method reduces the number of FLOPs with respect to block FD by a factor of $24.81$.

Regarding memory, the encoder for IDSE-RDO has complexity  $O(n_s n_p)$ since it stores $\numrma$ vectors of the size of the image. Although the values of $n_s$ we consider make the memory overhead manageable, future work may explore options to further improve memory complexity (\autoref{sec:conclusion}). 

\section{Empirical evaluation}
\label{sec:exper}

We test object detection/instance segmentation in the COCO 2017 validation set \cite{lin_microsoft_2014} and the PennFudan dataset \cite{wang_object_2007}. \newtext{In \autoref{sec:feat_reg}, \autoref{sec:large_scale}, and \autoref{sec:runtime}, we focus on testing the performance of different CV models and configurations using baseline AVC \cite{wiegand_overview_2003}. We apply IDSE-RDO to the luminance channel; for the chroma channels, we use SSE-RDO, setting a $\mathrm{QP}$ offset of $+3$. The RDO decides block-partitioning ($4\times 4$ or $16\times 16$) and quantization step ($\Delta \mathrm{QP} = -4, -3, \hdots, 3, 4$), i.e., the effective quantization step is derived from $\mathrm{QP} + \Delta \mathrm{QP}$. We use CAVLC for entropy coding \cite{wiegand_overview_2003}.}

\newtext{\autoref{sec:diff_codecs} explores a more realistic setup with advanced codecs, such as AVC FRExt \cite{sullivan2006overview} and HEVC \cite{sullivan_overview_2012}, providing a more complete assessment of the system performance. 
Since many edge devices are power-constrained, codecs such as AVC and HEVC remain relevant in the remote inference scenarios we address in this paper \cite{compressai_vision} even though higher RD performance codecs, e.g., VVC, are available.}


\newtext{As distortion metrics,} we report 1) mean average precision (mAP@[0.5:0.05:0.95]) \newtext{for both object detection and instance segmentation}, 2) Y-PSNR, and 3) Y-MS-SSIM \cite{wang_multiscale_2003}. \newtext{In \autoref{sec:diff_codecs} we also report the FD and the IDSE to illustrate the correlation of these metrics with accuracy in downstream tasks.} \newtext{Unless otherwise stated, }we set $\numrma=8$ as dimensionality after sketching, \newtext{which we found experimentally to yield a good trade-off between computational complexity and accuracy. Finding values of $n_s$ to guarantee a given degradation in downstream task performance as a function of the number of RDO candidates and the tolerance $\epsilon$ in \eqref{eq:samples} is left for future research}. By default, the downstream DNN is a Mask R-CNN with ResNet-50 \cite{he2017mask}, and we set $\tau = \tilde{\tau}$ from \eqref{eq:init_tau}. For simulations, we use an Intel(R) E5-2667 v4 with an NVIDIA GeForce RTX 3090 (24GB VRAM).

We evaluate four different feature extractors. Three of them are based on Mask R-CNN \cite{he2017mask}: a coarse output (P5) of an FPN \cite{lin_feature_2017}, which we call \fpnfe{}, and the outputs of the RPN, which we call RPN-FE. For RPN-FE, we will consider a Mask R-CNN with ResNet-50 (\rpnfefif{}) and a Mask R-CNN with ResNet-101 (\rpnfehun{}) \cite{xie2017aggregated}. The last feature extractor is the backbone of YOLOv9 \cite{wang2024yolov9} (\yolofe{}).

We explore the effect of the feature extractor and regularization (\autoref{sec:feat_reg}), as well as the coding performance in different tasks (\autoref{sec:large_scale}). Finally, we discuss complexity (\autoref{sec:runtime}) \newtext{and test more complex codecs (\autoref{sec:diff_codecs})}.

\subsection{Feature extractor and regularization}
\label{sec:feat_reg}
We consider $1000$ images from COCO   \cite{lin_microsoft_2014} and compress them with quality parameter $\mathrm{QP}\in\lbrace 27, 30, 33, 36, 39\rbrace$. \newtext{These QP values guarantee that the PSNR of the compressed images will lie in the interval $[30, 36]$ dB, a typical range of PSNRs that ensures noise in the image will be barely noticeable for a human observer \cite{wang_multiscale_2003}. In \autoref{sec:diff_codecs}, we will consider a wider range of QP values to explore the limits of the high bit-rate assumptions in \autoref{sec:taylor} and \autoref{sec:blockwise_localization}}

\subsubsection{Importance maps} 
\begin{figure*}
    \centering
    \includegraphics[width=\linewidth]{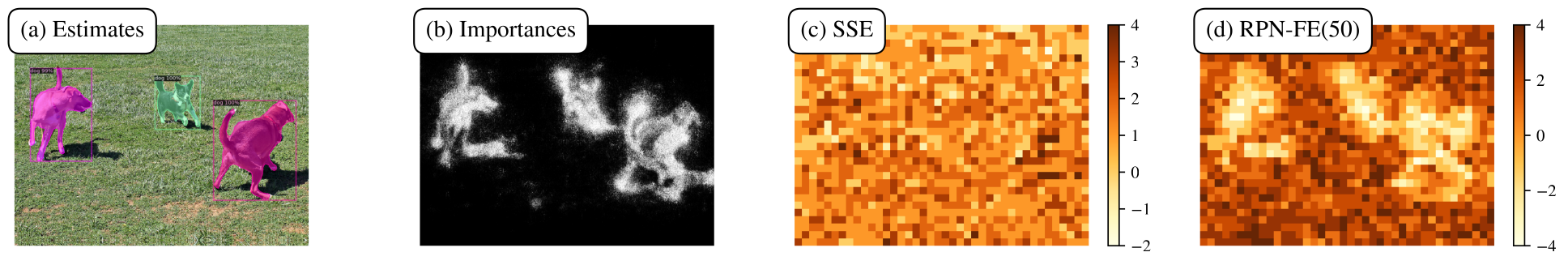}
    \caption{(a) Mask R-CNN estimates; (b) importances for \rpnfefif{}; (c-d) $\Delta\mathrm{QP}$ chosen by RDO with (c) SSE and (d) IDSE–\rpnfefif{}. Lower $\Delta\mathrm{QP}$ implies finer quantization. IDSE-RDO preserves relevant regions for the downstream task.}
    \label{fig:dQP}
\end{figure*}

\newtext{The Hessian definition in \eqref{eq:hessian} suggests that the squared norm of each row of the sketched Jacobian, $\mathrm{diag}(\jacos(\img)^\top \jacos(\img))$, yields an importance map, quantifying how much each region in an image affects a given downstream task.} 
To assess the effect of using different feature extractors, we depict pixel importance maps in \autoref{fig:semantic}. The feature extractors for \rpnfefif{} and \rpnfehun{} are deeper and discriminate better the regions in the image that are important for the target task. Our method applies Taylor's expansion first and then localizes the metric. An alternative is to localize the metric block-wise first, as in block FD \cite{fischer_video_2020}, and then apply Taylor’s expansion to each block as in \autoref{sec:taylor}. However, as \autoref{fig:semantic}(b) shows, evaluating the feature extractor block-wise leads to estimates of the importance that do not match exactly the position of the objects. We also depict the $\Delta\mathrm{QP}$ chosen block-wise for compressing an image with $\mathrm{QP}=29$ using SSE-RDO and IDSE-RDO with \rpnfefif{} (\autoref{fig:dQP}). IDSE-RDO allocates more bit-rate to relevant regions for the downstream task.

 \subsubsection{Different models}
\label{sec:diff_models} We consider three downstream DNNs: Mask R-CNN with ResNet-50 and ResNet-101 \cite{xie2017aggregated}, and YOLOv9 \cite{wang2024yolov9}, which is \newtext{a simpler model with lower computational complexity and faster runtime} and might be better suited for low-cost edge devices \cite{choi_high_2018} (cf.~\autoref{sec:runtime}). For IDSE-RDO, we use \rpnfefif{}, \rpnfehun{}, and \yolofe{}, which correspond to the early stages of the three downstream DNNs we use, and then evaluate each of these systems using the compressed images. We show the BD rate savings \cite{bjontegaard_calculation_2001} with respect to SSE-RDO \cite{bjontegaard_calculation_2001}  for each possible configuration in \autoref{tab:feat_extract_empir}. We conclude that IDSE-RDO provides coding gains for all the downstream DNNs we consider, regardless of the feature extractor.

\begin{table*}[t]
    \setlength{\tabcolsep}{9pt}
    \renewcommand{\arraystretch}{1.1}    
    \centering
    \begin{tabular}{@{}lccccccccc@{}}
        \toprule
         \multirow{3}{*}{$f(\cdot)$} & \multicolumn{2}{c}{\textbf{Visual quality}} & \multicolumn{4}{c}{\textbf{Mask R-CNN}} & \multicolumn{2}{c}{\textbf{YOLOv9}} \\
         \cmidrule(lr){2-3} \cmidrule(lr){4-7} \cmidrule(lr){8-9}
         & \multirow{2}{*}{\textbf{PSNR} ↓} & \multirow{2}{*}{\textbf{MS-SSIM} ↓} & \multicolumn{2}{c}{\textbf{ResNet-50}} & \multicolumn{2}{c}{\textbf{ResNet-101}} & \multirow{2}{*}{\textbf{mAP det.} ↓} & \multirow{2}{*}{\textbf{mAP seg.} ↓} \\
         &  &  & \textbf{mAP det.} ↓ & \textbf{mAP seg.} ↓ & \textbf{mAP det.} ↓ & \textbf{mAP seg.} ↓ & & \\
        \midrule
            \rpnfefif{} & $2.05$ & $2.32$ & $-8.65$ & $-9.30$  & $-8.26$ & $-8.60$ & $-7.47$ & $-6.45$ \\	
            \rpnfehun{} & \textcolor{customblue}{$\mathbf{1.92}$} & $2.44$ & \textcolor{customblue}{$\mathbf{-9.63}$} & \textcolor{customblue}{$\mathbf{-9.65}$} &\textcolor{customblue}{$\mathbf{-8.45}$} & \textcolor{customblue}{$\mathbf{-9.14}$} & \textcolor{customblue}{$\mathbf{-7.50}$} & \textcolor{customblue}{$\mathbf{-6.55}$}  \\	
            \yolofe{} & $2.53$ & \textcolor{customblue}{$\mathbf{1.96}$} & $-6.92$ & $-7.38$ & $-5.85$ & $-6.16$ & $-5.34$ & $-6.00$  \\	
        \bottomrule
    \end{tabular}    
    \caption{BD-rate savings [$\%$] with respect to SSE-RDO using IDSE-RDO. Lower is better ↓. We compute mAP for object detection (mAP det.) and instance segmentation (mAP seg.) using Mask R-CNN with ResNet-50 and ResNet-101, as well as YOLOv9. The best results for each column appear in \textcolor{customblue}{\textbf{blue}}. IDSE-RDO outperforms SSE-RDO for any configuration.}
    \label{tab:feat_extract_empir}
\end{table*}

\subsection{Assessing different tasks}
\label{sec:large_scale}
\subsubsection{COCO and transfer learning}
\begin{figure}
    \centering
    \includegraphics[width=\linewidth]{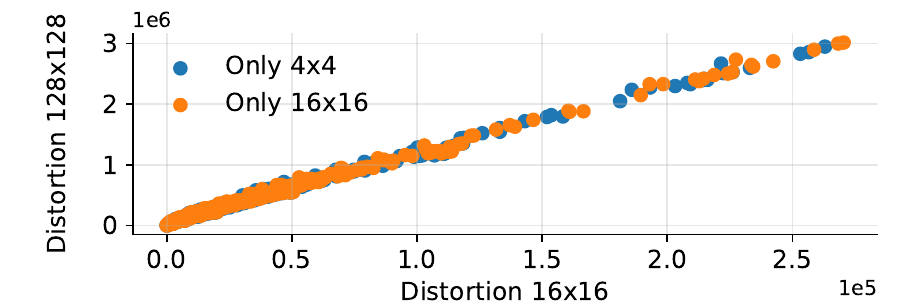}
	\caption{Block FD with blocks of  $128\times 128$ pixels (original \cite{fischer_video_2020}) and the corresponding $64$ sub-blocks of $16\times 16$ (approximation). To remove the effect of RDO, we use only $4\times 4$ or $16\times 16$ block partitions. The Pearson correlation coefficient is $0.997$ for both setups.}    
    \label{fig:128vs16}
\end{figure}

We use IDSE-RDO with \fpnfe{} and \rpnfefif{}. We also consider RDO with block FD in a setup inspired by \cite{fischer_video_2020}, using as a distortion metric the average of the block FD in the $5$th layer of VGG and the SSE. 
However, \cite{fischer_video_2020} used block sizes of $128\times 128$ pixels while, due to codec and resolution constraints, we use blocks of size $16\times 16$ pixels. 
To assess this approximation, we evaluate the feature distance using blocks of $128\times 128$ pixels (the original metric \cite{fischer_video_2020}) and the aggregate of the $64$ sub-blocks of $16\times 16$ pixels (our approximation). Correlation results (\autoref{fig:128vs16}) suggest the approximation is reasonable. As proposed in \cite{fischer_video_2020}, we implement block FD using the sum of absolute differences between the features rather than the square difference, and we set $c = 0.57$ in \eqref{eq:og_multiplier}, which also yields the best results. As in \cite{fischer_video_2020}, we re-scale the Lagrangian based on the ratio between SSE and block FD for the first block.

\begin{figure*}
    \centering
    \includegraphics[width=0.49\linewidth]{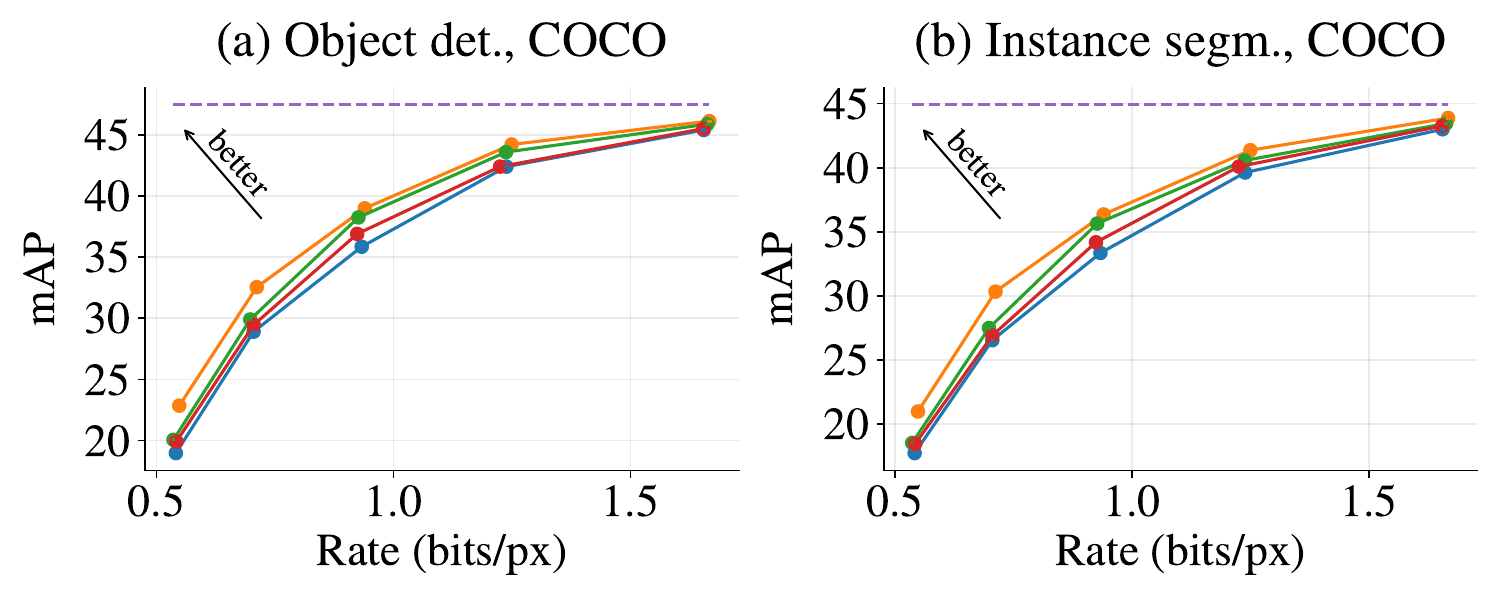}    
    \includegraphics[width=0.49\linewidth]{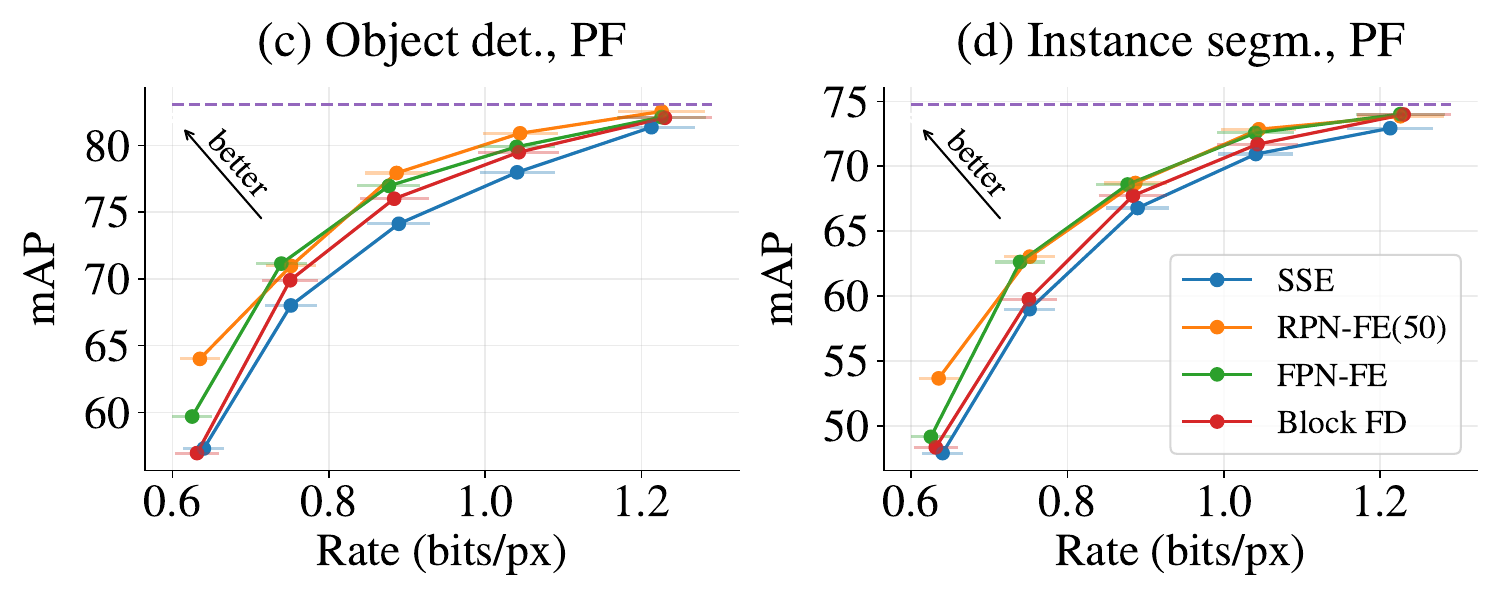}
    \caption{Rate-distortion (RD) curves for object detection and instance segmentation mAP using baseline AVC with SSE-RDO, IDSE-RDO with \fpnfe{} and \rpnfefif{}, and RDO using block FD on images from the COCO dataset (a--b) and the PennFudan dataset (c--d). We added the standard error on the estimation of the average bit-rate as a horizontal bar.}
    \label{fig:coco_pen}
\end{figure*}

Beyond images from COCO, we also follow a transfer learning approach, where we use the FPN from Mask R-CNN as a first step, and then we fine-tune the last layers to detect and segment pedestrians. For fine-tuning, we freeze the feature extractor and train the region proposal layers for $5$ epochs using a training set of $50$ images. We use the remaining $50$ images for testing. In this case, we compress with $\mathrm{QP}\in\lbrace 31, 33, 35, 37, 39\rbrace$ because task performance saturates for smaller $\mathrm{QP}$. RD curves (\autoref{fig:coco_pen}) and BD-rates (\autoref{tab:large_scale}) for COCO and PennFudan show that, for a similar performance in PSNR, IDSE-RDO yields coding gains with respect to block FD for mAP. Using \rpnfefif{} provides better mAP performance than \fpnfe{} since the former incorporates more information about the target task. 

\begingroup
	\begin{table}[t]
		\centering
        \renewcommand{\arraystretch}{1}    
		\setlength{\tabcolsep}{6pt} 
		\begin{tabular}{llccccc}
			\toprule & $f(\cdot)$ & \textbf{PSNR ↓} & \textbf{MS-SSIM ↓}
			& \textbf{Det. ↓} & \textbf{Seg. ↓} \\
			\midrule
   
            \multirow{3}{*}{\rotatebox{90}{\textbf{COCO}}}
			& \fpnfe{} & $\textcolor{customblue}{\mathbf{1.32}}$ & $\textcolor{customblue}{\mathbf{-3.36}}$ & $-6.06$ & $-5.93$ \\	
    		& \rpnfefif{} & $2.05$ & $2.32$ & $\textcolor{customblue}{\mathbf{-8.65}}$ & $\textcolor{customblue}{\mathbf{-9.30}}$ \\
            &Block FD & $1.84$ & $-0.91$ & $-2.18$ & $-2.23$ \\	
			\midrule
            \midrule
            \multirow{3}{*}{\rotatebox{90}{\textbf{PF}}}
			& \fpnfe{} & \textcolor{customblue}{$\mathbf{0.42}$} & \textcolor{customblue}{$\mathbf{-4.74}$} & $-7.31$ & $-6.00$ \\	
    		& \rpnfefif{} & $0.64$ & $0.94$ & \textcolor{customblue}{$\mathbf{-9.50}$} & \textcolor{customblue}{$\mathbf{-6.87}$} \\
            &Block FD & $0.68$ & $-1.26$ & $-4.25$ & $-3.21$ \\	
    		\bottomrule
    		\end{tabular}		
		\caption{BD-rate savings [\%] with respect to SSE-RDO. Lower is better ↓. The best method for each metric appears in \textcolor{customblue}{\textbf{blue}}. Our methods outperform SSE-RDO and block FD-RDO. \rpnfefif{} performs the best in CV tasks.}
		\label{tab:large_scale} 
	\end{table}
 \endgroup

\subsubsection{Mismatched models and self-supervised learning}
\label{sec:SSL_comp}
When the feature extractor does not match the downstream DNN, we expect our system to underperform. In this scenario, in which our information about the task is limited, we can rely on a feature extractor trained by SSL using augmentations that encode generic properties, such as rotational invariance. To test this idea, we construct two datasets from the COCO2017 validation set: a people dataset and a fruit dataset, each comprising $200$ images. We fine-tune a Mask R-CNN with a ResNet-50 to each dataset using subsets of $100$ images. We use these two systems as our downstream DNNs. 

We compress images with three \fpnfe{}: the FPNs from the two fine-tuned Mask R-CNNs above, and an additional FPN trained via SSL \cite{bardes2021vicreg} using geometric augmentations, which should be suitable for both person/fruit detection. We show the performance for each \fpnfe{} in \autoref{tab:SSL_results} for the remaining $100$ images of each dataset. The fine-tuned \fpnfe s reach the best results for the tasks they were trained for, while the SSL-based \fpnfe{} is the second best in both cases. If the feature extractor does not match the downstream DNN, we still have coding gains, but the performance drops.

\begin{table}[t]
    \centering
    \setlength{\tabcolsep}{4.5pt}
    \renewcommand{\arraystretch}{1.1}
    \begin{tabular}{@{}lcccccc@{}}
        \toprule
        \multirow{2}{*}{\textbf{$f(\cdot)$}} 
        & \multirow{2}{*}{\textbf{PSNR ↓}} 
        & \multirow{2}{*}{\textbf{MS-SSIM ↓}} 
        & \multicolumn{2}{c}{\textbf{People}} 
        & \multicolumn{2}{c}{\textbf{Fruits}} \\
        \addlinespace[0.2em] 
        & & & \textbf{Det.} ↓ & \textbf{Seg.} ↓ & \textbf{Det.} ↓ & \textbf{Seg.} ↓ \\
        \midrule      
        SSL   & \textcolor{customblue}{$\mathbf{0.85}$} & \textcolor{customblue}{$\mathbf{-3.67}$} & \textcolor{customred}{$\mathbf{-3.01}$} & \textcolor{customred}{$\mathbf{-3.08}$} & \textcolor{customred}{$\mathbf{-4.27}$} & \textcolor{customred}{$\mathbf{-4.11}$} \\  
        SL-People  & $1.04$ & \textcolor{customred}{$\mathbf{-2.90}$}  & \textcolor{customblue}{$\mathbf{-5.33}$} & \textcolor{customblue}{$\mathbf{-4.62}$} & $-3.01$ & $-3.04$ \\  
        SL-Fruits  & \textcolor{customred}{$\mathbf{0.91}$} & $-2.44$  & $-2.20$ & $-1.94$ & \textcolor{customblue}{$\mathbf{-6.61}$} & \textcolor{customblue}{$\mathbf{-4.32}$} \\              
        \bottomrule
    \end{tabular}
    \caption{BD-rate savings [\%] with respect to SSE-RDO for people detection/segmentation and fruit detection/segmentation. SL stands for supervised learning. Lower is better ↓. The best method for each metric appears in \textcolor{customblue}{\textbf{blue}}, the second best in \textcolor{customred}{\textbf{orange}}.}
    \label{tab:SSL_results} 
\end{table}

\subsection{Computational complexity}
\label{sec:runtime}
\autoref{tab:times} shows the average runtime to compute the Jacobian for $100$ images of the COCO dataset \cite{lin_microsoft_2014} for \yolofe{}, \rpnfefif{}, and \rpnfehun{}. 
The computational complexity scales linearly with the number of samples used to sketch the Jacobian. 
We compute the runtime of \newtext{sketching the} Jacobian and encoding the image using IDSE-RDO, and we measure the increase with respect to the runtime of using SSE-RDO. The worst case overhead is $11.02\%$, which compares favorably with using block FD \cite{fischer_video_2020} for RDO  (\autoref{sec:large_scale}), which results in $89\%$ computational overhead. \newtext{Next, we analyze the performance with other codecs.}

\begingroup
	\begin{table}[t]
		\centering
        \renewcommand{\arraystretch}{1.2}
        \setlength{\tabcolsep}{6pt} 
		\begin{tabular}{llcccc}
			\toprule & $f(\cdot)$
			&  $\numrma = 2$  & $\numrma = 4$ & $\numrma = 8$ & $\numrma = 16$ \\
			\midrule
   
            \multirow{3}{*}{\rotatebox{90}{\makecell{\textbf{Compute}\\\textbf{Jacobian}}}} &
            \rpnfefif{} &  \textcolor{customred}{$\mathbf{0.097}$} &  \textcolor{customred}{$\mathbf{0.162}$} &  \textcolor{customred}{$\mathbf{0.292}$} &  \textcolor{customred}{$\mathbf{0.557}$}\\	
            
            & \rpnfehun{} & {$0.123$} & {$0.206$} & {$0.368$} & {$0.699$}\\
            & \yolofe{} & \textcolor{customblue}{$\mathbf{0.063}$} & \textcolor{customblue}{ $\mathbf{0.097}$} & \textcolor{customblue}{ $\mathbf{0.168}$} & \textcolor{customblue}{ $\mathbf{0.307}$} \\
            \midrule
            \multirow{3}{*}{\rotatebox{90}{\makecell{\textbf{Encoder}\\\textbf{overhead}}}} &
			\rpnfefif{} &  \textcolor{customred}{$\mathbf{2.67\%}$} &  \textcolor{customred}{$\mathbf{4.13\%}$} &  \textcolor{customred}{$\mathbf{7.24\%}$} &  \textcolor{customred}{$\mathbf{11.55\%}$}\\	
            & \rpnfehun{} &$3.22 \%$ & $4.94\%$ & $7.39 \%$ & $12.02 \%$\\
            & \yolofe{} &\textcolor{customblue}{$\mathbf{2.04 \%}$} &\textcolor{customblue}{$\mathbf{3.72 \%}$} &\textcolor{customblue}{$\mathbf{7.15 \%}$} & \textcolor{customblue}{$\mathbf{10.27 \%}$} \\
            \bottomrule
    		\end{tabular}		
\caption{Average runtime [s] to compute $\jacos(\img)$ for different $n_s$ (top) and increase in encoder runtime [$\%$] with respect to SSE-RDO (bottom). In both cases, lower is better. The best values for each column appear in \textcolor{customblue}{\textbf{blue}}, the second best in \textcolor{customred}{\textbf{orange}}. \yolofe{} is the most efficient, followed by \rpnfefif{} and \rpnfehun{}.}
\label{tab:times}
    \end{table}	 
\endgroup

\subsection{Different codecs}
\label{sec:diff_codecs}
\newtext{We evaluate our method on two other standardized codecs: AVC FRExt (JM-19.1) \cite{sullivan2006overview} and HEVC (HM-18.0) \cite{sullivan_overview_2012}. Both use the main profile, and we apply the same quantization level to luma and chroma channels. For AVC FRExt, IDSE-RDO is used to select block partitions ($4\times 4$, $8\times 8$, $16\times16$), trellis quantization (search space of $\pm 1$), and the $16\times16$ intra-prediction mode. For HEVC, we use IDSE-RDO for block partition and trellis quantization but SSE-RDO for transform type selection, as we found it had minimal impact on downstream task performance. We hypothesize that transform choice mainly adapts to residual statistics rather than redistributing the bit-rate based on semantic properties. Importantly, our framework supports any RDO-driven decision; this setup simply exemplifies a practical use case.}

\subsubsection{AVC FRExt}
\begin{figure*}
    \centering
    \includegraphics[width=\linewidth]{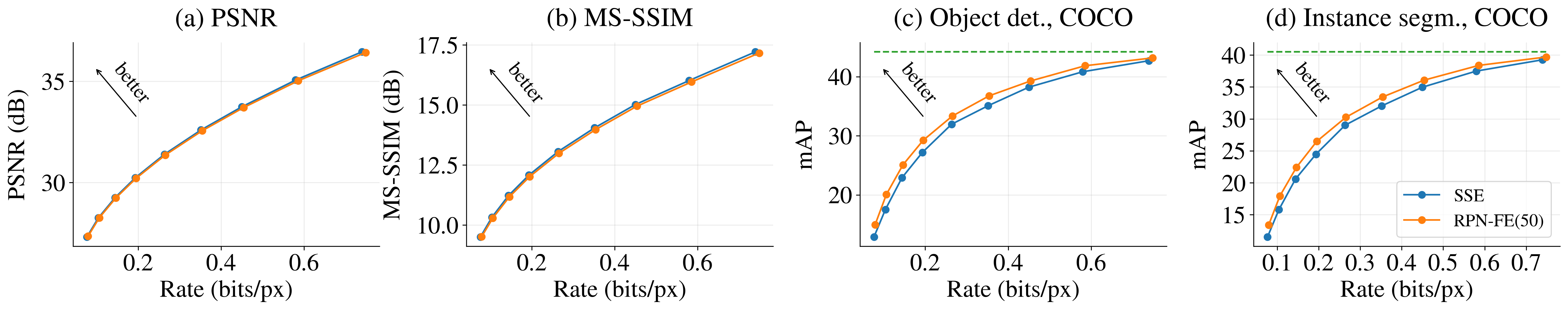}
    \caption{\newtext{RD curves for PSNR, MS-SSIM, object detection, and instance segmentation mAP using AVC FRExt with SSE-RDO and our proposed IDSE-RDO with \rpnfefif{} on $5000$ images from the COCO dataset. We added the standard error on the estimation of the average bit-rate as a horizontal bar. IDSE-RDO outperforms SSE-RDO even at lower bit-rates.}}
    \label{fig:results_avc_frext}
\end{figure*}

\newtext{We first evaluate AVC FRExt on 5000 COCO images using SSE-RDO and IDSE-RDO with \rpnfefif{}. We consider QP between $31$ and $47$ in steps of $2$. As shown in \autoref{fig:results_avc_frext}, IDSE-RDO consistently outperforms SSE-RDO in downstream task accuracy while introducing some penalty in PSNR and MS-SSIM. These results suggest that our high bit-rate approximations in \eqref{eq:jaco} and \eqref{eq:localization} remain accurate even at the coarsest quantization steps we consider.}

\begin{figure}
    \centering
    \includegraphics[width=\linewidth]{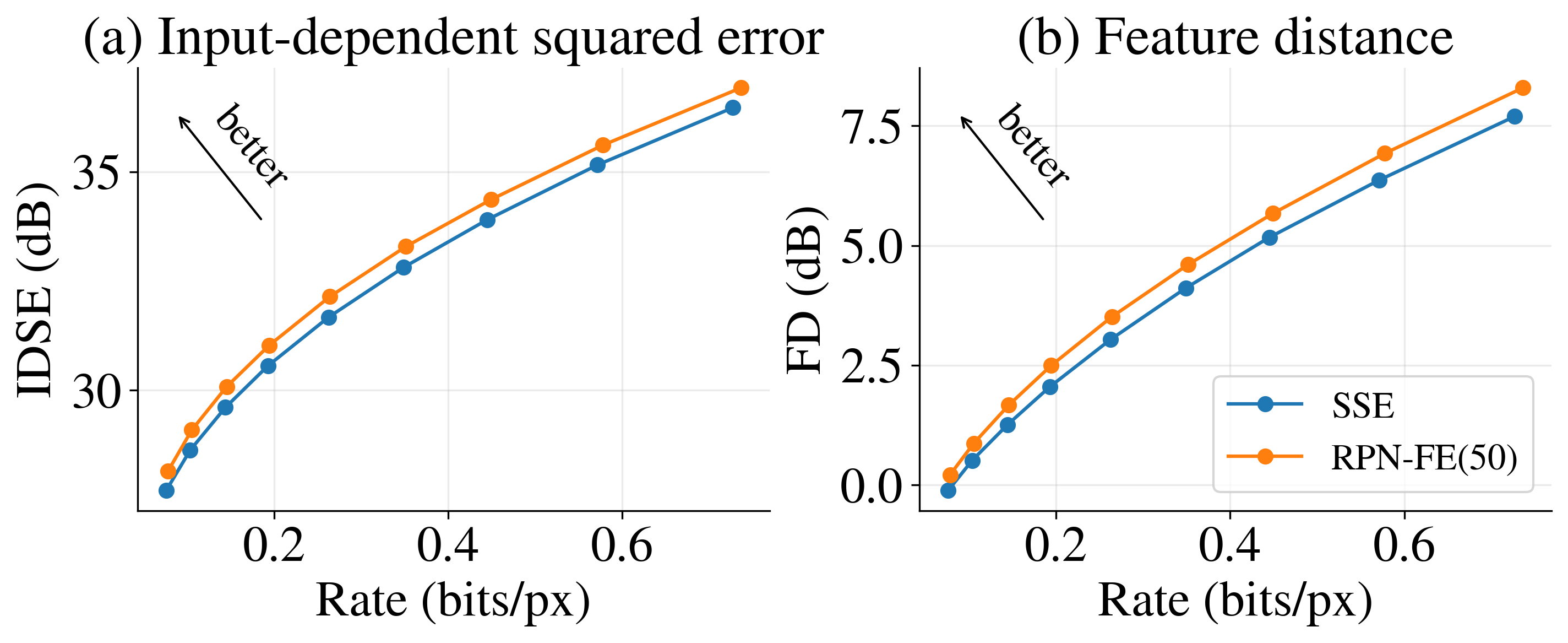}
    \caption{\newtext{RD curves for FD and IDSE, both of them with \rpnfefif{}, averaged across $1000$ images from the COCO dataset, using AVC FRExt with SSE-RDO and IDSE-RDO with \rpnfefif{}. Our proposed IDSE-RDO also provides coding gains when considering these two metrics, demonstrating correlation with accuracy in downstream tasks.}}
    \label{fig:rd_fd_idse}
\end{figure}

\begin{figure}
    \centering
    \includegraphics[width=1\linewidth]{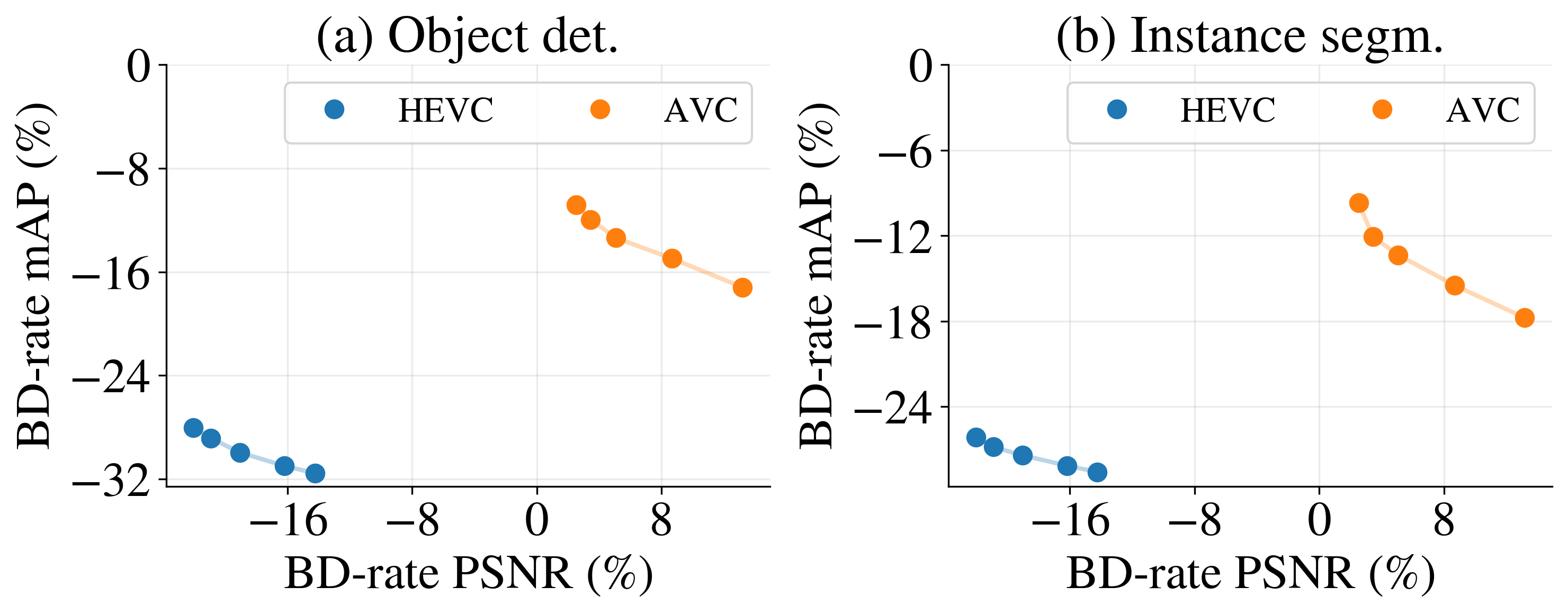}
    \caption{\newtext{BD-rate PSNR vs BD-rate mAP for object detection and instance segmentation in $1000$ images from the COCO dataset, using IDSE-RDO for multiple values of $\tau$ in AVC FRExt and HEVC. The baseline is SSE-RDO AVC FRExt. We can improve mAP in exchange for some PSNR degradation.}}
    \label{fig:sweep_frext}
\end{figure}

\newtext{To assess the alignment between IDSE, FD, and mAP, we report RD curves using FD and IDSE as distortion metrics (\autoref{fig:rd_fd_idse}) and provide BD-rate values in \autoref{tab:multiple_metrics}. IDSE-RDO improves performance across all metrics, reinforcing the link between FD, IDSE, and task accuracy.} \newtext{We also explore the trade-off between task accuracy and perceptual quality by sweeping $\alpha\in\lbrace 0.25, 0.50, 1.00, 1.50, 2.00\rbrace$ in $\tau = \alpha \tilde{\tau}$ \eqref{eq:distortion_balance}. As shown in \autoref{fig:sweep_frext}, increasing $\alpha$ yields up to 17\% BD-rate gain in mAP for object detection and instance segmentation at the cost of a 12\% rate increase for PSNR. These trade-offs can be tuned to fit specific applications.}

\begin{table}[t]
    \centering
    \setlength{\tabcolsep}{7pt} 
    \renewcommand{\arraystretch}{1}
    \begin{tabular}{@{}lcccc}
        \toprule
         \textbf{Codec} & \textbf{FD ↓} & \textbf{IDSE ↓} &  \textbf{Det. ↓} & \textbf{Seg. ↓} \\
        \midrule
        
            AVC/FRExt, IDSE & $-10.60$ & $-10.35$ & $-10.69$ & $-10.69$ \\
			HEVC, SSE & $-17.02$ & $-14.57$ & ${-19.76}$ & ${-17.76}$ \\
			HEVC, IDSE & $\textcolor{customblue}{\mathbf{-28.32}}$ & $\textcolor{customblue}{\mathbf{-24.84}}$ & $\textcolor{customblue}{\mathbf{-29.10}}$ & $\textcolor{customblue}{\mathbf{-26.19}}$ \\       
        \bottomrule
    \end{tabular}    
    \caption{\newtext{BD-rate savings [$\%$] with respect to SSE-RDO AVC FRExt for $1000$ images in COCO considering FD, IDSE, and mAP in object detection and instance segmentation. Lower is better ↓. For IDSE-RDO, we use \rpnfefif{}. The best value for each column appears in \textcolor{customblue}{\textbf{blue}}. Improvements in IDSE and FD correlate with improvements in mAP.}}
    \label{tab:multiple_metrics}
\end{table}	
\newtext{Next, we analyze the trade-off between computational complexity and coding performance by varying $n_s$ (cf.~\autoref{tab:times}). Results in \autoref{tab:num_samples} show that increasing $n_s$ improves both PSNR and mAP. Notably, even with $n_s=2$, IDSE-RDO achieves substantial gains—exceeding 12.5\% BD-rate improvement over SSE-RDO.}

\begin{table}[t]
    \centering
    \setlength{\tabcolsep}{7pt} 
    \renewcommand{\arraystretch}{1}
		\begin{tabular}{llccccc}
			\toprule & \textbf{Sketch.} & \textbf{PSNR ↓} & \textbf{MS-SSIM ↓}
			& \textbf{Det. ↓} & \textbf{Seg. ↓} \\
			\midrule
   
            \multirow{4}{*}{\rotatebox{90}{\textbf{AVC}}}
			& $n_s=2$ & $5.78$ & $4.95$ & $-12.65$ & $-12.72$ \\	
    		& $n_s=4$ & $5.31$ & $4.49$ & $-13.86$ & $-12.92$ \\
            &$n_s=8$ & $5.05$ & $4.34$ & $-13.94$ & $-13.09$ \\
            &$n_s=16$ & $\textcolor{customblue}{\mathbf{4.91}}$ & $\textcolor{customblue}{\mathbf{4.21}}$ & $\textcolor{customblue}{\mathbf{-15.05}}$ & $\textcolor{customblue}{\mathbf{-15.00}}$ \\	
			\midrule
            \midrule
            \multirow{4}{*}{\rotatebox{90}{\textbf{HEVC}}}
			& $n_s=2$ & $5.12$ & $4.68$ & $-9.28$ & $-10.16$ \\	
    		& $n_s=4$ & $4.95$ & $4.32$ & $-10.46$ & $-10.88$ \\	
            &$n_s=8$ & $4.67$ & $4.11$ & $-11.71$ & $-11.93$ \\	
            &$n_s=16$ & $\textcolor{customblue}{\mathbf{4.42}}$ & $\textcolor{customblue}{\mathbf{3.92}}$ & $\textcolor{customblue}{\mathbf{-12.32}}$ & $\textcolor{customblue}{\mathbf{-12.61}}$ \\	
    		\bottomrule
    		\end{tabular}		
    \caption{\newtext{BD-rate savings [$\%$] for IDSE-RDO with AVC FRExt and HEVC. The baseline is the SSE-RDO version of the corresponding codec. Lower is better ↓. For IDSE-RDO, we use \rpnfefif{} with $\tau = \tilde{\tau}/2$ in \eqref{eq:init_tau}. The best values for each column appear in \textcolor{customblue}{\textbf{blue}}. Increasing the number of sketching samples improves the performance of our method.}}
    \label{tab:num_samples}
\end{table}	

\subsubsection{HEVC}
\begin{figure}
    \centering
    \includegraphics[width=\linewidth]{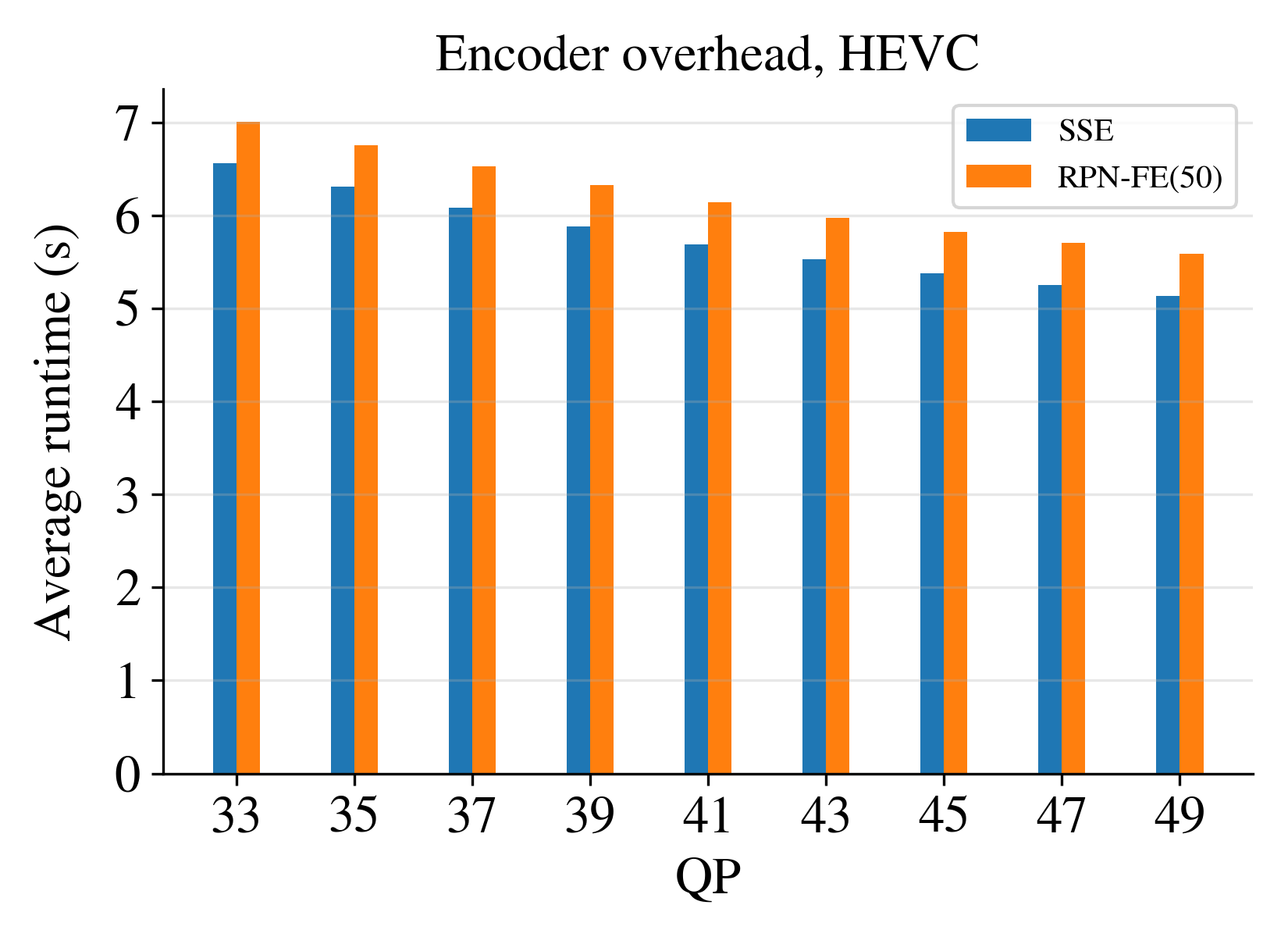}
    \caption{\newtext{Average runtime over $200$ images from the COCO dataset for the HEVC encoder using SSE-RDO and IDSE-RDO with \rpnfefif{}, as a function of the QP value. The average encoder overhead is $7.86\%$.}}
    \label{fig:complexity_increase}
\end{figure}

\newtext{We evaluate HEVC using both SSE-RDO and IDSE-RDO on COCO. We set QP $\in\{31, 35, 39, 43, 47\}$. As shown in \autoref{tab:multiple_metrics}, IDSE-RDO yields significant improvements in FD, IDSE, and downstream accuracy compared to HEVC with SSE-RDO and AVC FRExt.} \newtext{We also evaluate the impact of IDSE-RDO on encoder runtime. \autoref{fig:complexity_increase} shows the average runtime over 200 COCO images as a function of QP, including the runtime for computing the Jacobian. Using \(n_s = 8\), the overhead relative to SSE-RDO is only 7.86\%, emphasizing the practicality of our method.}

\newtext{Finally, we assess the trade-off between BD-rate for mAP and BD-rate for PSNR with HEVC by sweeping $\alpha\in\lbrace 0.25, 0.50, 1.00, 1.50, 2.00\rbrace$ in $\alpha\tilde{\tau}$ (\autoref{fig:sweep_frext}). While IDSE-RDO allows similar trade-offs as in AVC, the range of achievable gains is smaller. We conjecture that IDSE-RDO is applied to some of the RDO steps, but many RDO decisions in HEVC are still based on SSE–e.g., the choice of intra-prediction angle–limiting the ability to redistribute bit-rate based on IDSE. Future work will explore the impact of different RDO choices on downstream task performance. Another possible reason is that, with HEVC, we are closer to the best mAP (e.g., performance based on the uncompressed images). The higher rate points for IDSE-RDO HEVC with $\tau = \tilde{\tau}$, corresponding to $0.61 \ \mathrm{and} \ 0.47 \mathrm{ \ bits/px}$ achieve 42.1\% and 41.9\% for object detection and 39.8\% and 39.5\% for instance segmentation, which is close to the $42.4\%$ for object detection and $40.2\%$ for instance segmentation we obtain with uncompressed images. }

\section{Conclusion \newtext{and future work}}
\label{sec:conclusion}
In this paper, we proposed an RDO method that preserves the feature distance (FD). Using linearization arguments and a localization assumption, we simplified the FD to an input-dependent squared error (IDSE) loss involving the Jacobian of the feature extractor. To make the metric computable, we proposed a sketching method for the Jacobian. IDSE can be computed block-wise and in the transform domain. We validated our method using \newtext{AVC and HEVC, showing coding gains of up to $17\%$ for AVC and $12\%$ for HEVC} for computer vision tasks \newtext{with a $7.86\%$ encoder complexity increase.} 

Future work will address memory consumption. For instance, storing the importance maps instead of the sketched Jacobian has memory complexity $O(\numpix)$. We could improve upon this result via vector quantization of the importance map \cite{fernandez_image_2023}. Other extensions include IDSE for feature compression \cite{choi_deep_2018} and transform design \cite{fernandez2024fast}, as well as \newtext{considering} other codecs such as VVC \cite{bross2021overview}. Although we expect \newtext{similar} results, \newtext{codecs with a larger number of RDO options (e.g., multiple transform set selection) might require higher dimensional sketches, increasing memory and runtime overhead.} 

\bibliographystyle{IEEEbib}
\bibliography{IEEEabrv,original_submission}

\makeatletter
\AddToHook{enddocument/afteraux}{%
  \immediate\write18{cp output.aux first_review.aux}%
}
\makeatother
\end{document}